\documentclass[twocolumn]{aastex631}
\newcommand\teff{\mbox{$T_\mathrm{eff}$}}
\newcommand\vtan{\mbox{$V_\mathrm{tan}$}}
\newcommand\mjup{\mbox{$M_\mathrm{Jup}$}}
\newcommand\kms{{km s$^{-1}$}}
\usepackage{graphicx, color} 

\begin{document}

\title{New Ultracool Companions to Nearby White Dwarfs}

\author[0009-0002-3936-8059]{Alexia Bravo}
\affil{United States Naval Observatory, Flagstaff Station, 10391 West Naval Observatory Rd., Flagstaff, AZ 86005, USA}

\author[0000-0002-6294-5937]{Adam C.~Schneider}
\affil{United States Naval Observatory, Flagstaff Station, 10391 West Naval Observatory Rd., Flagstaff, AZ 86005, USA}

\author[0000-0003-2478-0120]{Sarah Casewell}
\affil{School of Physics and Astronomy, University of Leicester, University Road, Leicester, LE1 7RH, UK}

\author[0000-0003-4083-9962]{Austin Rothermich}
\affil{Department of Astrophysics, American Museum of Natural History, Central Park West at 79th St., New York, NY 10024, USA}

\author[0000-0001-6251-0573]{Jacqueline K. Faherty}
\affil{Department of Astrophysics, American Museum of Natural History, Central Park West at 79th St., New York, NY 10024, USA}

\author[0000-0003-0825-4876]{Jenni R. French}
\affil{School of Physics and Astronomy, University of Birmingham, Edgbaston, Birmingham, B15 2TT, UK}

\author[0000-0003-2235-761X]{Thomas P. Bickle}
\affil {School of Physical Sciences, The Open University, Milton Keynes, MK7 6AA, UK}
\affil {Backyard Worlds: Planet 9}

\author[0000-0002-1125-7384]{Aaron M. Meisner}
\affil{NSF's National Optical-Infrared Astronomy Research Laboratory, 950 N. Cherry Ave., Tucson, AZ 85719, USA}

\author[0000-0003-4269-260X]{J. Davy Kirkpatrick}
\affil{IPAC, Mail Code 100-22, Caltech, 1200 E. California Blvd., Pasadena, CA 91125, USA}

\author[0000-0002-2387-5489]{Marc J. Kuchner}
\affil{Exoplanets and Stellar Astrophysics Laboratory, NASA Goddard Space Flight Center, 8800 Greenbelt Road, Greenbelt, MD 20771, USA}

\author[0000-0002-6523-9536]{Adam J. Burgasser}
\affil{Center for Astrophysics and Space Science, University of California San Diego, La Jolla, CA 92093, USA}

\author[0000-0001-7519-1700]{Federico Marocco}
\affil{IPAC, Mail Code 100-22, Caltech, 1200 E. California Blvd., Pasadena, CA 91125, USA}

\author[0000-0002-1783-8817]{John H. Debes}
\affil{AURA for ESA, Space Telescope Science Institute, 3700 San Martin Drive, Baltimore, MD 21218, USA}

\author[0000-0003-4864-5484]{Arttu Sainio}
\affil{Backyard Worlds: Planet 9}

\author[0000-0002-8960-4964]{L\'eopold Gramaize}
\affil{Backyard Worlds: Planet 9}

\author[0000-0001-8662-1622]{Frank Kiwy}
\affil{Backyard Worlds: Planet 9}

\author[0000-0002-4175-295X]{Peter A. Jałowiczor}
\affil{Backyard Worlds: Planet 9}

\author[0009-0009-9014-7535]{Awab Abdullahi}
\affil{Backyard Worlds: Planet 9}

\begin{abstract}
We conducted a search for new ultracool companions to nearby white dwarfs using multiple methods, including the analysis of colors and examination of images in both the optical and the infrared. Through this process, we identified fifty-one previously unrecognized systems with candidate ultracool companions. Thirty-one of these systems are resolved in at least one catalog, and all but six are confirmed as co-moving companions via common proper motion and consistent parallax measurements (when available). We have followed up four co-moving companions with near-infrared spectroscopy and confirm their ultracool nature. The remaining twenty candidates are unresolved, but show clear signs of infrared excess which is most likely due to the presence of a cold, low-mass companion or a dusty circumstellar disk. Three of these unresolved systems have existing optical spectra that clearly show the presence of a cool stellar companion to the white dwarf primary via spectral decomposition. These new discoveries, along with our age estimates for the primary white dwarfs, will serve as valuable benchmark systems for future characterization of ultracool dwarfs.

\end{abstract}

\section{Introduction} 
The first L-dwarf (GD 165B) was discovered as a companion to a nearby white dwarf \citep{becklin1988}. While thousands of L, T, and Y type dwarfs have been discovered since, the number of confirmed substellar companions to white dwarf primaries has remained relatively small. Other L, T, and Y type companions to white dwarfs include PHL 5038B \citep{steele2009}, LSPM J1459+0857B \citep{dayjones2011}, WD 0806-661B \citep{luhman2011}, Wolf 1130C \citep{mace2013}, LSPM J0241+2553B \citep{deacon2014}, LSPM J0055 + 5948AB \citep{meisner2020}, COCONUTS-1B \citep{zhang2020}, SDSS J222551.65+001637.7B \citep{french2023}, and VVV 1256$-$62AB \citep{zhang2024}. Several more candidates have also recently been uncovered (e.g., \citealt{kiwy2022, rothermich2024}) and are awaiting spectroscopic confirmation. 

Despite their relative rarity, each system containing a white dwarf with a resolved ultracool dwarf (spectral type later than $\sim$M7) companion holds great value as a benchmark system. It is typically very difficult to determine a precise age of a solitary ultracool dwarf, and often their distances are difficult to measure because of their inherent faintness. If a cool companion can be physically tied to a white dwarf primary, the age and distance of the white dwarf, which are often much easier to determine, can be applied to the cold secondary.

For objects with masses below the substellar limit (brown dwarfs), companionship can be especially valuable because brown dwarfs cool over time, resulting in a degenerate relationship between age, effective temperature (\teff), and mass. White dwarf primaries with measurable ages can break this degeneracy leading to much more precise physical property determination for any brown dwarf companion (e.g., \citealt{luhman2011, mace2018, zhang2020}). White dwarfs in particular have very predictable cooling rates, which is why they are often used to determine the ages of nearby stars (e.g., \citealt{fouesneau2019, qui2021, baig2024}), stellar clusters (e.g, \citealt{salaris2018, lodieu2019a, lodieu2019b, heyl2022}), and larger populations (e.g., \citealt{kilic2019}).

Additional low-mass companions to white dwarfs have been found through indirect means, such as radial velocity variations (e.g., \citealt{maxted2006, casewell2012, farihi2017}), photometric transits (e.g., \citealt{beuermann2013, parsons2017, casewell2020b}), or photometric/spectral decomposition (e.g., \citealt{farihi2004, dobbie2005, steele2011, steele2013, casewell2018, casewell2020, owens2023}).  These systems, however, are often products of post-common envelope evolution.  While their utility as age benchmarks is not as clear as with wide ultracool companions, these systems provide rare laboratories for studying the effects of radiation on substellar atmospheres. 

We have used the the substantial increase in the number of known nearby ($<$100 pc) white dwarfs enabled by Gaia \citep{gaia2021} in \cite{gentile2021} to search for new ultracool companions in the solar neighborhood. We describe our candidate selection methods in Section \ref{sec:cans}, follow-up observations in Section \ref{sec:obs}, system properties in Section \ref{sec:sys}, an analysis of  newly discovered systems in Section \ref{sec:analysis}, and provide a conclusion in Section \ref{sec:conclusion}.

\section{Candidate Search}
\label{sec:cans}

\begin{figure*}
\plotone{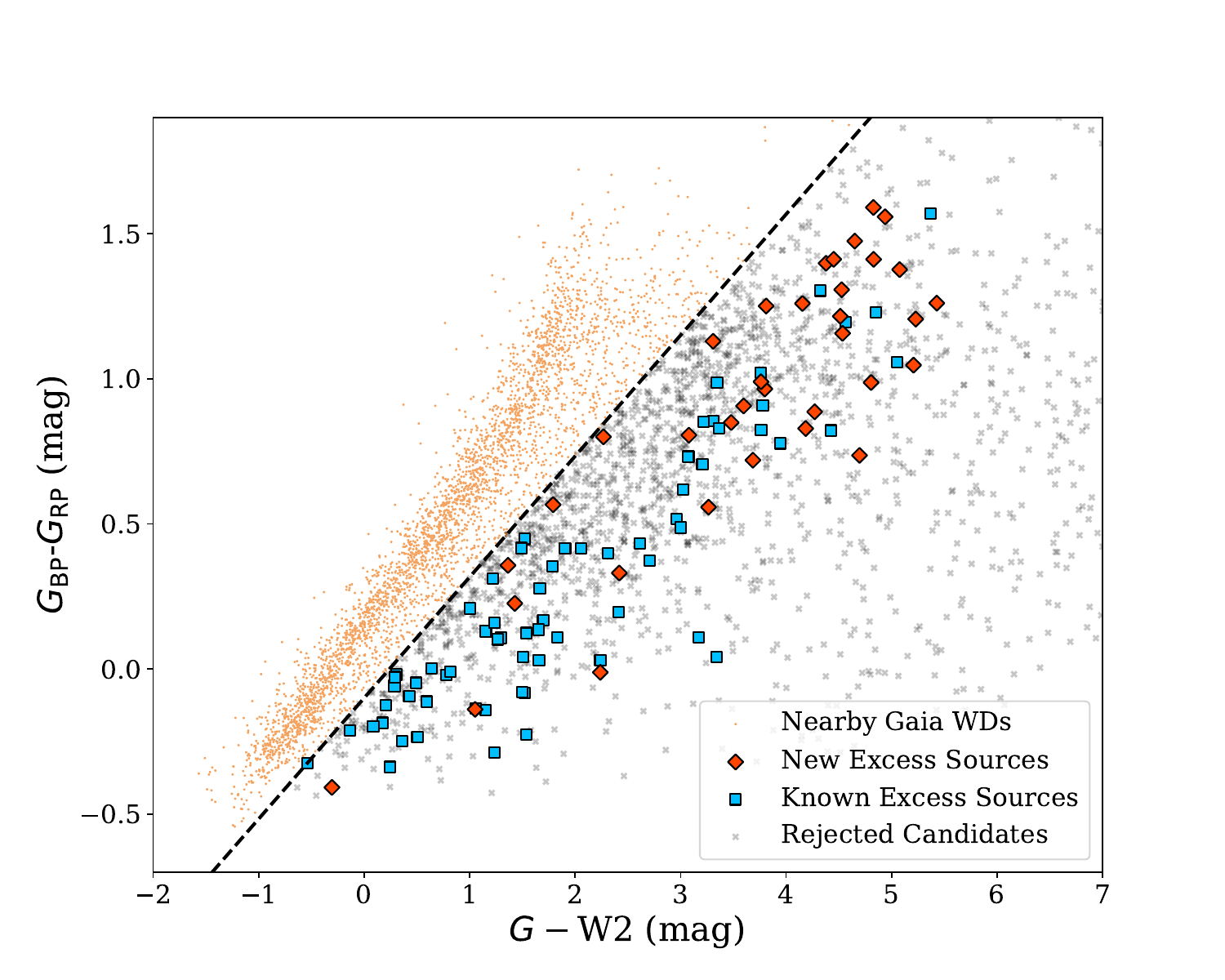}
\caption{$G_{\rm BP}$$-$$G_{\rm RP}$ vs.\ $G-$W2 colors for Gaia white dwarfs within 100 pc from \cite{gentile2021}. Objects on the normal white dwarf color sequence are displayed as orange dots. Objects showing signs of infrared excess were selected as having redder $G-$W2 colors than the dashed line. The color positions of previously known excess sources and newly discovered excess candidates selected via Method I are shown as blue squares and red diamonds, respectively. Rejected excess candidates are shown with grey x symbols.} 
\label{fig:plot1}
\end{figure*}

\subsection{The White Dwarf Sample}
Our initial list of white dwarfs came from \cite{gentile2021} which is derived from the Gaia Early Data Release 3 \citep{gaia2021} catalog and contains $\sim$1.3 million likely white dwarfs. Our objective is to search for ultracool companions, which are inherently faint. We therefore limited our initial sample to white dwarfs within 100 pc ($\varpi$ $\geq$ 10 mas). This distance requirement resulted in a candidate list of 16,959 objects.

We employed three different search methods with this sample in order to identify a comprehensive set of both unresolved and resolved candidate companions. These three methods included identifying color outliers, visually inspecting optical and infrared images of each white dwarf for cool companions, and finding ultracool candidates using Gaia astrometry and photometry. Each method is described in detail in the following three sections. 

\subsubsection{Method I: Color Cuts}
\label{sec:mthd1}

This method involved identifying color-selected outliers within our sample of white dwarf candidates. We cross-matched each source with the nearest detection in the CatWISE2020 catalogue \citep{marocco2021}.  We propagate the positions of each candidate white dwarf to the CatWISE 2020 epoch ($\sim$2015) using Gaia proper motions and use a 5\arcsec\ search radius.  We use the proper motion corrected CatWISE 2020 photometry (e.g., w1mpro\_pm and w2mpro\_pm) for all sources.  We also cross-matched each white dwarf with available near-infrared surveys using proper motion corrected coordinates, including the UKIRT Hemisphere Survey (UHS; \citealt{dye2018}); the UKIRT Infrared Deep Sky Survey (UKIDSS) Large Area Survey (LAS; \citealt{lawrence2007}), Galactic Plane Survey (GPS; \citealt{lucas2008}), and Galactic Clusters Survey (GCS; \citealt{lawrence2007}); the Visible and Infrared Survey Telescope for Astronomy Hemisphere Survey (VHS; \citealt{mcmahon2013}), VISTA Variables in the Via Lactea (VVV; \citealt{minniti2010}), VISTA Magellanic Clouds (VMC; \citealt{cioni2011}), and VISTA Kilo-degree Infrared Galaxy (VIKING; \citealt{edge2013}) surveys. We then created several color-color plots using Gaia, near-infrared, and CatWISE magnitudes. 

Using a sample known white dwarf-brown dwarf systems from the literature (including both resolved and unresolved companions), we searched for color combinations that differentiated the known white dwarf-brown dwarf sample from the rest of the white dwarf population. We determined that the $G_{\rm BP}$$-$$G_{\rm RP}$ vs.\ $G-$W2 color color diagram clearly distinguished these two populations. The Gaia filters have pivot wavelengths of 621.79 nm, 510.97 nm, and 776.91 nm for $G$, $G_{\rm BP}$, and $G_{\rm RP}$, respectively \citep{riello2018}, while the WISE W2 filter has a reference wavelength of 4.6028 $\mu$m \citep{wright2010}.  The non-excess white dwarfs show a clear correlation between $G_{\rm BP}$$-$$G_{\rm RP}$ and $G-$W2 colors, with redder $G_{\rm BP}$$-$$G_{\rm RP}$ colors corresponding to redder $G-$W2 colors.  Both of these colors have been shown to correlate with absolute $G-$band magnitudes, with redder colors corresponding to fainter absolute magnitudes (e.g., \citealt{gaia2018, kirkpatrick2024}).  Thus this trend can generally be viewed as a temperature sequence, with the coldest white dwarfs having the reddest $G_{\rm BP}$$-$$G_{\rm RP}$ and $G-$W2 colors.  We selected as candidates those objects with $G_{\rm BP}$$-$$G_{\rm RP}$ colors less than 0.417$\times$($G-$W2) $-$ 0.10 mag (see Figure \ref{fig:plot1}), which cleanly selected a large sample of known, blended white dwarf/ultracool dwarf systems. However, this color selection criteria still suffered from significant contamination, mostly from blending with bright sources with small separations. We attempted to eliminate spurious candidates by first flagging those sources with another Gaia source within 3\arcsec, thereby eliminating many white dwarf candidates that were blended with a nearby, bright Gaia source. We also required CatWISE2020 W2 magnitudes to have a signal-to-noise ratio (S/N) greater than 10. 

These cuts resulted in a total of 3,095 candidates. Each of these candidates was then individually inspected using available optical and infrared images. The vast majority of these sources still showed clear signs of contamination, typically in the form of blending with an unrelated nearby source. The remaining sources, highlighted in Figure \ref{fig:plot1}, either showed no clear contamination, or were found to have a resolved companion within a few arcseconds. 

\subsubsection{Method II: Visual Inspection}
\label{sec:mthd2}
This method entailed the manual inspection of available optical images, such as those from the Panoramic Survey Telescope and Rapid Response System (Pan-STARRS (PS1); \citealt{chambers2016}) and the Dark Energy Survey (DES; \citealt{abbott2021}). For each of the 16,959 objects in our white dwarf sample, we inspected optical images (when available) to identify by-eye any sources within $\sim$30\arcsec\ with exceptionally red optical colors. A search radius of 30\arcsec\ was chosen because it is sufficiently large to expand the search area covered by Method I, but is also small enough to not be subject to an excessive amount of false positives.  While this search recovered some of the same sources found using the method in Section \ref{sec:mthd1} that were resolved in optical or near-infrared images, but not resolved in the Wide Field Infrared Survey Explorer (WISE; \citealt{wright2010}) imagery, it was also sensitive to wider separation companions that would not easily be identified by colors of the white dwarf alone.    

Red colors do not guarantee that a candidate companion is co-moving with the white dwarf primary. For those with Gaia parallaxes and proper motions, we ensured consistency between candidate companions and the nearby white dwarfs. For those candidate companions not in Gaia, we either used proper motions from relevant catalogs, such as CatWISE2020 or the NOIRLab Source Catalog DR2 (NSC; \citealt{nidever2021}), or calculated proper motions from available catalog detections from UKIDSS, UHS, or PS1. For UHS and UKIDSS, positions were calibrated using the Gaia DR3 reference frame following \cite{schneider2023}.  For PS1, detections were taken directly from the DR2 catalog.  If detections existed in UKIRT and PS1 catalogs, they were not mixed to avoid systematics between catalogs. Time baselines were several years from each catalog, and thus parallactic motion should have a negligible effect on the resulting proper motions. We ensured that companion candidates had proper motions and parallax (when available) measurements that agreed to within 5$\sigma$, or had proper motion measurements with differences less than 5 mas yr$^{-1}$ and parallax measurement differences less than 1 mas.  The second selection criteria was necessary for pairs with exceptionally small measurement uncertainties.  For example, while the A and B components of the 4\arcsec\ separation WD J1046+6054AB system have $\mu_{\alpha}$ and $\mu_{\delta}$ differences of only 3.2 and 2.7 mas yr$^{-1}$, respectively, their small uncertainties result in 18$\sigma$ and 13$\sigma$ disagreements.  

There are two exceptions to the above requirements.  The first is the companion candidate to WD J001237.06$-$492422.81, which initially had proper motion differences of less than 2$\sigma$ using the relatively imprecise proper motion measurement of the companion candidate from CatWISE 2020 \citep{marocco2021}.  A more precise proper motion measurement was later found in NSC DR2 \citep{nidever2021}, resulting in larger discrepancies between $\mu_{\alpha}$ (14$\sigma$) and $\mu_{\delta}$ (4$\sigma$) components.  While this object is still listed in Table \ref{tab:resastro}, it is unlikely to be a physical pair.  Second, the putative companion to WD J211901.60$+$420617.34 was observed spectroscopically before being fully vetted kinematically.  We include it in Table \ref{tab:resastro} despite its unlikely physical nature.  For a few candidates, we were unable to produce a proper motion precise enough to clearly establish companionship. We retain these sources for completeness, but label them as ``unreliable'' until more precise proper motions can be measured. 

We also estimated spectral types for each candidate companion. The focus of this work is to identify new companions with spectral types $\sim$M7 and later. We used Gaia parallaxes of candidate companions when available, or the parallax of the putative white dwarf primaries, to calculate absolute $G$-, $J$- and $K$-band magnitudes. We compared these values to the absolute $G$-band magnitude versus spectral type table from \cite{kiman2019}, absolute $J$- and $K$-band values from  \cite{pecaut2013}, and the absolute $J$- and $K$-band versus spectral type relations from \cite{schneider2023} to estimate spectral types for each companion. Estimated spectral types from different passbands generally agreed to within $\pm$1 subtype.  In cases where the different relations resulted in different spectral type estimates, the median type was chosen.  Candidates with estimated types of $\sim$M7 and later were retained.  

During the visual examination of optical images for objects from \cite{gentile2021}, we identified two objects that appeared unlikely to be white dwarfs based on their red-optical colors (WD J224600.88$-$060947.02 and WD J235432.47$+$080404.64). WD J224600.88$-$060947.02 was previously identified by \cite{obrien2024} as a likely brown dwarf contaminant in the \cite{gentile2021} sample.  We suggest WD J235432.47$+$080404.64 is also likely a low-mass star or brown dwarf. Using Gaia parallaxes and photometry from VHS for WD J224600.88$-$060947.02 and UKIDSS LAS for WD J235432.47$+$080404.64, we find absolute magnitude consistent with spectral types of $\sim$L2 for WD J224600.88$-$060947.02 and $\sim$L1 for WD J235432.47$+$080404.64. 

\subsubsection{Method III: Gaia Comparisons}
\label{sec:mthd3}
Our final method of identifying potential ultracool companions used available Gaia astrometry and photometry for objects within the vicinity of each nearby white dwarf. We identified every object with Gaia distances $<$110 pc within 60\arcsec\ of each object in our white dwarf sample. We chose 110 pc to account for some uncertainty in both white dwarf and candidate companion parallax measurements. For each object within 60\arcsec\ of one of our white dwarf targets that had a distance measurement within 110 pc, we individually checked for relative proper motion and parallax consistency, to be further evaluated with the \texttt{CoMover} tool (see Section \ref{sec:rescans}). We further selected for possible ultracool companions by identifying those sources with absolute $G$-band magnitudes $>$14.5 mag, which corresponds to a spectral type of $\sim$M8. We did not require any further color-cuts based on Gaia colors, as the redness of ultracool sources can make the Gaia photometry unreliable \citep{smart2019}. However, we did note that some white dwarfs were found to match to each other (i.e., white dwarf-white dwarf systems). These candidate systems are listed in Appendix \ref{sec:appA}.

\subsection{Final Candidates}

\subsubsection{Unresolved Candidates}
\label{sec:unres}
\begin{figure*}
\plotone{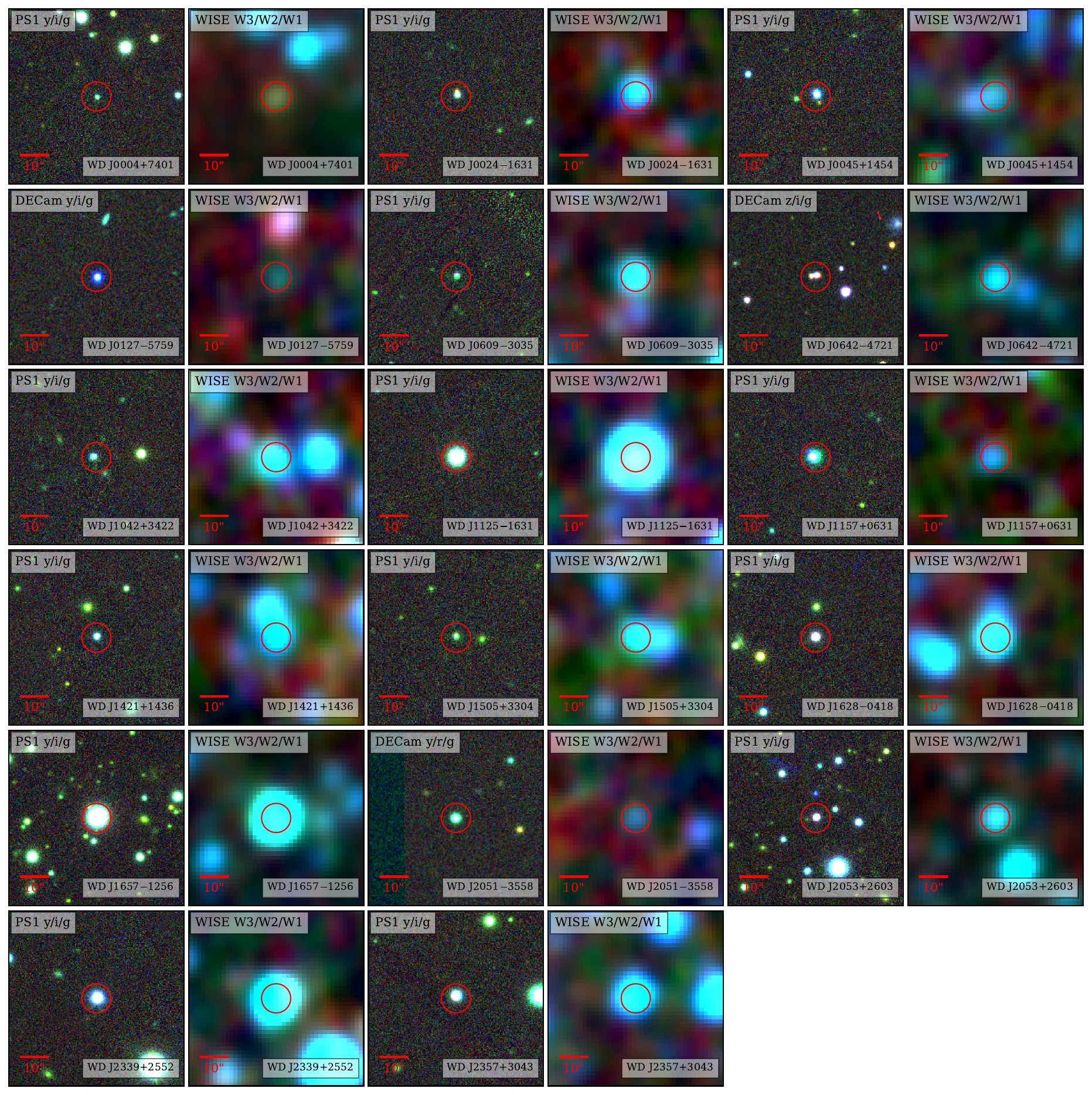}
\caption{Three color RGB optical and WISE images of all new unresolved candidates. Optical images either come from Pan-STARRS (PS1; \citealt{chambers2016}) or DECam images obtained through the Astro Data Lab image cutout service \citep{fitzpatrick2014, nikutta2020}. WD identifiers are listed for each object. } 
\label{fig:cutouts2a}
\end{figure*}

Our unresolved candidates were found solely through Method I (Section \ref{sec:mthd1}). A total of seventy-four unresolved candidate systems were identified. We searched the literature for each candidate to determine if these objects were previously known to have infrared excess, which could be indicative of an unresolved companion or dusty debris disk. Fifty-seven of our unresolved candidates were determined to have been previously identified as having either a known debris disk, an unresolved companion, or general infrared excess. These sources are listed in Table \ref{tab:unresrecovered}. Seventeen of our unresolved candidates had not been previously reported. These objects are listed in Table \ref{tab:unresolved} with their corresponding Gaia and CatWISE2020 photometric measurements. Optical and WISE images for each of these candidates is shown in Figure \ref{fig:cutouts2a}. The infrared excess of these white dwarfs could suggest the presence of a surrounding dust disk, or alternatively an unresolved low-mass companion (explored further in Section \ref{sec:analysis}).

We note that the results of our search for white dwarfs showing signs of excess infrared emission are different from previous searches because of our specific color-selection criteria and the construction of our input sample.  Because we are most interested in new, cold companions to white dwarfs, a distance-limited sample is most appropriate, as such companions are inherently faint.  Other infrared-excess searches focused on white dwarfs (e.g., \citealt{favieres2024}) instead use a magnitude limited sample, which would be sensitive to excess emission around bright white dwarfs beyond 100 pc that are not included in our sample. 

\startlongtable
\begin{deluxetable}{lc}
\tabletypesize{\footnotesize}
\tablecaption{Previously Known IR Excess Systems}
\label{tab:unresrecovered}
\tablehead{\colhead{System} & \colhead{Ref}} 
\startdata
WD J004912.03$+$384130.49 & 1 \\
WD J005045.82$-$032655.47 & 1 \\
WD J005815.51$-$563810.16 & 2 \\
WD J010726.22$+$251835.60 & 1 \\
WD J010749.38$+$210745.84 & 1 \\
WD J010933.16$-$190117.56 & 3 \\
WD J011221.14$-$561427.51 & 4 \\
WD J013326.41$+$040104.57 & 5 \\
WD J013532.98$+$144555.90 & 6 \\
WD J014754.82$+$233943.60 & 7 \\
WD J020524.94$-$794103.78 & 1 \\
WD J020733.81$+$333129.53 & 8 \\
WD J021912.81$-$630654.86 & 9 \\
WD J022320.55$-$045906.66 & 1 \\
WD J023427.73$-$045430.68 & 10 \\
WD J030253.10$-$010833.80 & 11 \\  
WD J033645.45$+$704411.16 & 1 \\
WD J035817.13$+$462839.74 & 12 \\  
WD J040435.02$+$150226.60 & 13 \\  
WD J040911.40$-$711741.56 & 14 \\  
WD J041653.28$+$262420.41 & 15 \\  
WD J041937.77$-$730344.51 & 16 \\  
WD J043354.57$+$282729.07 & 1 \\
WD J043839.37$+$410932.35 & 17 \\  
WD J051002.15$+$231541.42 & 1 \\
WD J053112.43$-$245120.86 & 2 \\
WD J054224.17$+$014414.64 & 18 \\  
WD J070245.87$+$000318.99 & 1 \\
WD J075508.95$-$144550.95 & 21 \\  
WD J080227.74$+$563155.46 & 1 \\
WD J081149.35$+$421208.95 & 18 \\  
WD J093341.31$-$100009.20 & 1 \\
WD J094755.68$-$231234.10 & 13 \\  
WD J100609.17$+$004417.08 & 19 \\  
WD J101728.52$-$323608.84 & 1 \\
WD J101803.83$+$155158.57 & 17 \\  
WD J105212.55$+$332318.39 & 1 \\
WD J111424.68$+$334124.39 & 20 \\  
WD J111912.40$+$022033.05 & 17 \\  
WD J121314.48$+$114050.14 & 2 \\ 
WD J122305.09$-$293228.29 & 3 \\
WD J125224.22$-$291456.00 & 22 \\  
WD J130542.40$+$180103.77 & 23 \\  
WD J131246.44$-$232132.60 & 24 \\  
WD J142833.74$+$440346.94 & 1 \\
WD J143830.09$-$312719.79 & 9 \\
WD J145006.66$+$405535.20 & 16 \\  
WD J154144.89$+$645352.98 & 18 \\  
WD J160839.55$+$172336.67 & 25 \\  
WD J161316.63$+$552125.97 & 18 \\  
WD J165126.05$+$663506.34 & 26 \\  
WD J172349.68$+$045847.54 & 1 \\
WD J172845.71$+$205341.68 & 1 \\
WD J173134.32$+$370520.71 & 27 \\  
WD J190319.56$+$603552.65 & 1 \\
WD J222958.08$+$302410.01 & 1 \\
WD J225726.12$+$075542.60 & 25 \\  
WD J232847.64$+$051454.24 & 28 \\
WD J233401.45$+$392140.87 & 2 \\  
WD J234036.64$-$370844.72 & 10 \\
\enddata
\tablerefs{(1) \cite{rebassa2019}; (2) \cite{favieres2024}; (3) \cite{dennihy2017}; (4) \cite{girven2012}; (5) \cite{ren2018}; (6) \cite{casewell2020}; (7) \cite{wang2019}; (8) \cite{debes2019}; (9) \cite{raddi2017}; (10) \cite{thorstensen2016}; (11) \cite{casewell2020}; (12) \cite{greenstein1986}; (13) \cite{xu2020}; (14) \cite{kraft1965}; (15) \cite{ren2014}; (16) \cite{hoard2013}; (17) \cite{farihi2005}; (18) \cite{girven2011}; (19) \cite{schultz1996}; (20) \cite{raymond2003}; (21) \cite{rebassa2010}; (22) \cite{mumford1969}; (23) \cite{nather1981}; (24) \cite{ruiz2001}; (25) \cite{debes2011}; (26) \cite{lai2021}; (27) \cite{kilic2005}; (28) \cite{zuckerman1987}; (29) \cite{rodriguez2005}}
\end{deluxetable}

\begin{longrotatetable}
\begin{deluxetable*}{lccccccccccc}
\tabletypesize{\footnotesize}
\tablecaption{Unresolved Candidates}
\label{tab:unresolved}
\tablehead{\colhead{} & \colhead{$\varpi$} & \colhead{$\mu_{\alpha}$} & \colhead{$\mu_{\delta}$} & \colhead{$G$} & \colhead{$G_{\rm BP}$} & \colhead{$G_{\rm RP}$} & \colhead{W1} & \colhead{W2} & \\
\colhead{WD Name} & \colhead{(mas)} & \colhead{(mas yr$^{-1}$)} & \colhead{(mas yr$^{-1}$)} & \colhead{(mag)} & \colhead{(mag)} & \colhead{(mag)} & \colhead{(mag)} & \colhead{(mag)} &}
\startdata
WD J000427.01$+$740141.77&10.8108$\pm$0.4736&25.751$\pm$0.600&92.855$\pm$0.587&20.2099$\pm$0.0060&20.7608$\pm$0.0843&19.3846$\pm$0.0558&16.214$\pm$0.025&15.133$\pm$0.025&\\
WD J002455.00$-$163113.24&12.7572$\pm$0.3538&91.026$\pm$0.383&-73.071$\pm$0.291&19.4282$\pm$0.0039&20.0190$\pm$0.0542&18.7119$\pm$0.0300&15.199$\pm$0.020&14.901$\pm$0.027&\\
WD J004558.03$+$145449.80&15.3367$\pm$0.1207&-9.667$\pm$0.159&-140.873$\pm$0.094&18.0909$\pm$0.0030&18.2887$\pm$0.0115&17.7224$\pm$0.0159&16.492$\pm$0.036&16.299$\pm$0.088&\\
WD J012709.06$-$575914.45&10.5455$\pm$0.0691&-125.975$\pm$0.074&-100.091$\pm$0.074&17.7421$\pm$0.0030&17.7279$\pm$0.0100&17.8669$\pm$0.0136&17.577$\pm$0.062&16.690$\pm$0.092&\\
WD J060942.14$-$303552.74&13.8173$\pm$0.2077&26.909$\pm$0.218&-151.904$\pm$0.251&19.2061$\pm$0.0037&19.6818$\pm$0.0413&18.4664$\pm$0.0269&14.936$\pm$0.017&14.692$\pm$0.021&\\
WD J064244.74$-$472110.25&11.7956$\pm$0.3079&-29.702$\pm$0.326&162.089$\pm$0.371&19.8052$\pm$0.0052&20.1959$\pm$0.0865&18.7842$\pm$0.0513&15.219$\pm$0.020&14.975$\pm$0.021&\\
WD J104203.04$+$342227.34&12.7616$\pm$0.3765&-289.601$\pm$0.414&-49.101$\pm$0.401&19.8001$\pm$0.0043&20.2390$\pm$0.0602&19.2517$\pm$0.0414&15.504$\pm$0.019&14.994$\pm$0.027&\\
WD J112542.43$-$163154.24&21.5638$\pm$0.0429&11.886$\pm$0.048&-2.11$\pm$0.037&15.8768$\pm$0.0028&16.1853$\pm$0.0036&15.2789$\pm$0.0042&12.495$\pm$0.012&12.279$\pm$0.009&\\
WD J115745.93$+$063148.28&13.0494$\pm$0.0877&-270.468$\pm$0.101&-24.079$\pm$0.065&17.2563$\pm$0.0029&17.3923$\pm$0.0067&17.0351$\pm$0.0081&16.308$\pm$0.034&15.892$\pm$0.064&\\
WD J142152.24$+$143646.09&10.4251$\pm$0.2389&14.538$\pm$0.270&-68.285$\pm$0.210&18.9430$\pm$0.0037&19.3551$\pm$0.0381&18.3900$\pm$0.0285&15.322$\pm$0.019&15.144$\pm$0.032&\\
WD J150549.34$+$330410.82&10.0124$\pm$0.4198&-22.958$\pm$0.299&-100.046$\pm$0.414&20.1376$\pm$0.0048&20.7498$\pm$0.0775&19.3516$\pm$0.0493&16.120$\pm$0.024&15.758$\pm$0.041&\\
WD J162816.20$-$041810.26&11.0206$\pm$0.1284&-55.89$\pm$0.135&-40.297$\pm$0.101&17.9737$\pm$0.0030&18.4794$\pm$0.0147&17.2198$\pm$0.0108&13.957$\pm$0.015&13.818$\pm$0.015&\\
WD J165737.30$-$125634.46&11.0374$\pm$0.0383&6.688$\pm$0.041&-10.134$\pm$0.031&15.0652$\pm$0.0029&15.3129$\pm$0.0033&14.5071$\pm$0.0044&12.178$\pm$0.013&11.986$\pm$0.009&\\
WD J205148.28$-$355820.33&11.9003$\pm$0.1023&23.587$\pm$0.089&-16.476$\pm$0.079&17.4747$\pm$0.0029&17.5565$\pm$0.0082&17.3303$\pm$0.0098&16.699$\pm$0.043&16.046$\pm$0.071&\\
WD J205357.06$+$260326.29&10.2420$\pm$0.1493&15.049$\pm$0.140&-23.11$\pm$0.138&18.5542$\pm$0.0031&19.1209$\pm$0.0279&17.7092$\pm$0.0135&14.286$\pm$0.014&14.103$\pm$0.015&\\
WD J233910.91$+$255205.81&11.1455$\pm$0.0623&12.292$\pm$0.054&-32.474$\pm$0.044&16.5273$\pm$0.0029&16.6828$\pm$0.0053&16.1250$\pm$0.0053&13.481$\pm$0.012&13.262$\pm$0.011&\\
WD J235745.75$+$304357.77&13.8718$\pm$0.5094&5.298$\pm$0.449&-203.983$\pm$0.330&17.7454$\pm$0.0041&17.8467$\pm$0.0100&16.8568$\pm$0.0068&14.199$\pm$0.013&13.983$\pm$0.014&\\
\enddata
\tablecomments{All positions and optical photometry come from Gaia \citep{gaia2021} and mid-infrared photometry comes from CatWISE 2020 \citep{marocco2021}.}
\end{deluxetable*}
\end{longrotatetable}

\subsubsection{Resolved Candidates}
\label{sec:rescans}
A combined total of 54 candidate white dwarf + ultracool dwarf systems that were resolved in at least one catalog were identified using Methods I, II, III. A literature search revealed that 23 of these systems had been discovered previously. These recovered systems are listed in Table \ref{tab:resrecovered}.

\begin{deluxetable}{lc}
\tabletypesize{\footnotesize}
\tablecaption{Known Resolved Systems}
\label{tab:resrecovered}
\tablehead{\colhead{System} & \colhead{Ref}} 
\startdata
WD J000801.26$-$350449.62ABC & 1 \\  
WD J000837.22$+$351144.52AB & 2 \\  
WD J013608.94$-$255614.10AB & 2 \\  
WD J015501.26$-$130749.19AB & 2 \\  
WD J024149.28$+$255344.60AB & 3 \\  
WD J035556.50$+$452510.26AB & 4 \\  
WD J052933.35$-$635655.29AB & 5 \\  
WD J064111.04$-$202743.83AB & 6 \\  
WD J082545.09$-$021024.73AB & 2 \\  
WD J095953.92$-$502717.75AB & 7 \\  
WD J120815.61$+$084543.16AB & 8 \\  
WD J123304.35$+$030245.83AB & 9 \\  
WD J124428.57$-$011857.85AB & 9 \\  
WD J125646.05$-$620208.54AB & 10 \\  
WD J131730.86$+$483333.05AB & 11 \\  
WD J155516.93$+$315306.96AB & 12 \\  
WD J162324.07$+$343647.37AB & 13 \\  
WD J164119.19$+$350425.08AB & 14 \\  
WD J184014.71$+$343846.50AB & 15 \\  
WD J185520.08$-$231440.90AB & 2 \\  
WD J192409.62$+$550648.48AB & 15 \\  
WD J221536.67$+$134711.17AB & 7 \\  
WD J234507.33$+$581315.07AB & 16 \\  
\enddata
\tablerefs{(1) \cite{tokovinin2022}; (2) Casewell et al.~(in prep.); (3) \cite{deacon2014}; (4) \cite{zhang2020}; (5) \cite{ravinet2024}; (6) \cite{girven2011}; (7) \cite{gcns2021}; (8) \cite{zhang2010}; (9) \cite{kiman2019}; (10) \cite{zhang2024}; (11) \cite{jalowiczor2021}; (12) \cite{kiman2021}; (13) \cite{silvestri2006}; (14) \cite{silvestri2007}; (15) \cite{lepine2005}; (16) \cite{rebassa2019}}
\end{deluxetable}

\begin{figure*}
\plotone{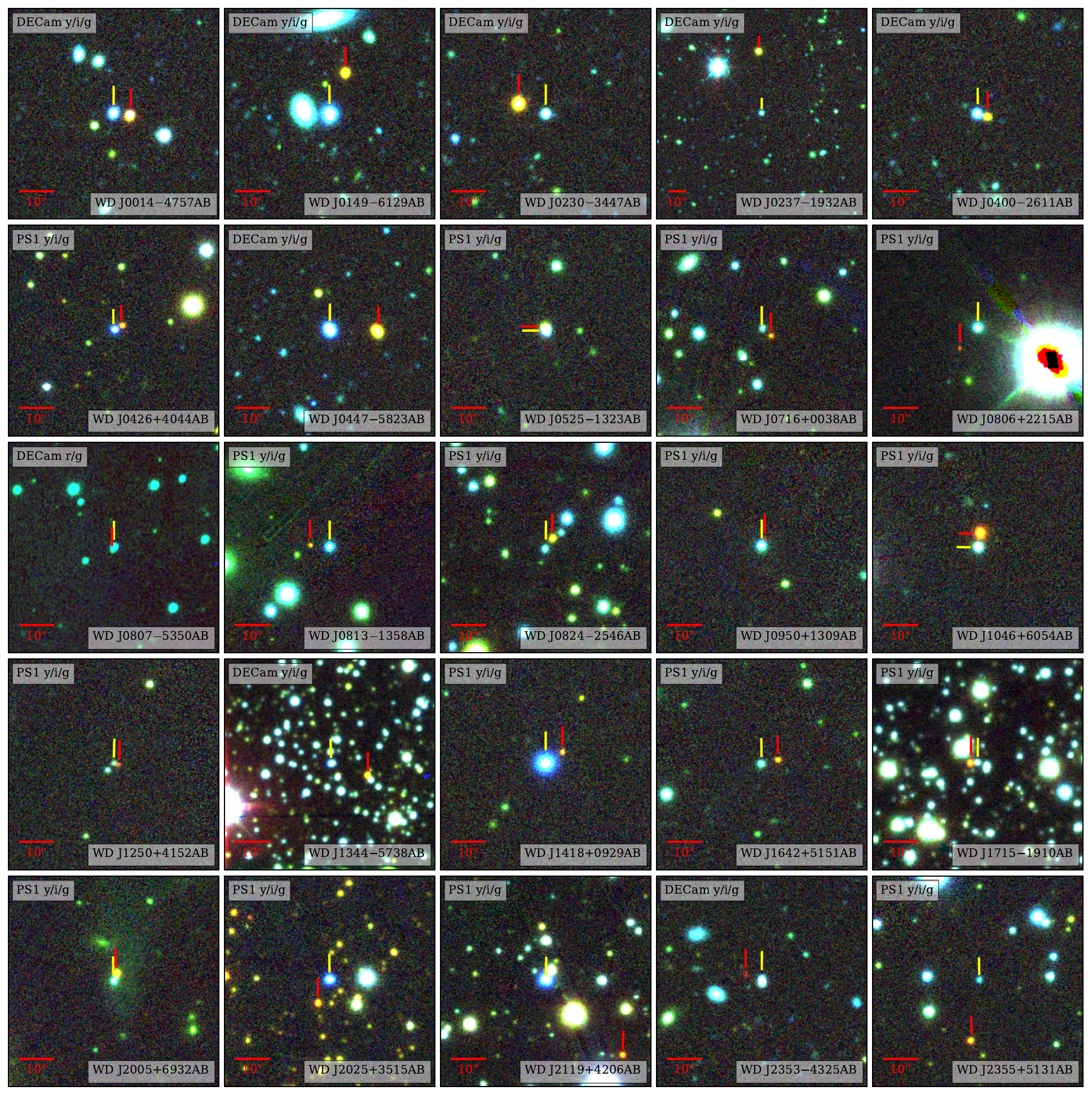}
\caption{Three color RGB images of all new resolved candidates. White dwarf primaries are marked with yellow lines while candidate ultracool companions are marked with red lines. The WD 0807$-$5350 system has only $r$- and $g$-band DECam images, and thus the candidate companion is difficult to see in the image (though is resolved in Gaia). Note that the ``unreliable'' candidates are not included in this figure.} 
\label{fig:cutouts1}
\end{figure*}

We found a total of 31 new systems with white dwarf primaries and candidate ultracool dwarf companions where the companions were resolved in at least one survey. Astrometric and photometric information are provided in Tables \ref{tab:resastro}, \ref{tab:resoptphoto}, and \ref{tab:resirphoto}. Available absolute magnitudes and spectral type estimates are given in Table \ref{tab:sptest}. Cutout optical images of the candidate systems are shown in Figure \ref{fig:cutouts1}. Note that the ``unreliable'' candidates are not included in Figure \ref{fig:cutouts1}.  All candidate systems were evaluated with the \texttt{CoMover} comoving probability tool, which determines the Bayesian probability of two sources being a physically comoving pair. \texttt{CoMover} probabilities are provided in Table \ref{tab:resastro}.   

Several of these resolved candidate companions had been noted as candidate ultracool dwarfs or high-proper motion objects in previous surveys, but had not been recognized as having a white dwarf companion. WD J014953.55$-$612916.97B, WD J023020.88$-$344749.38B, WD J044713.58$-$582318.04B, WD J104627.91$+$605448.32B, and WD J211901.60$+$420617.34B were each selected as Gaia ultracool dwarf candidates in \cite{reyle2018}. WD J044713.58$-$582318.04B was also previously identified as a high proper motion source in \cite{deacon2007} and \cite{kirkpatrick2016}. WD J104627.91$+$605448.32B has an optical spectral type of M9 in \cite{kiman2019}. WD J080648.05$+$221551.65B was selected as an L-type candidate object in \cite{galvez2017}. Lastly, WD J141835.29$+$092919.68B was previously identified as a photometric L dwarf candidate in \cite{skrzypek2016}. 

Several of these sources have also been identified as ultracool dwarf candidates by citizen scientists working with the Backyard Worlds: Planet 9 project \citep{kuchner2017}. We recognize the contributions of these scientists in the footnotes of Table \ref{tab:resastro}. 

\begin{longrotatetable}
\begin{deluxetable*}{ccccccccccccccc}
\tabletypesize{\footnotesize}
\tablecaption{Astrometry for Resolved Candidates}
\label{tab:resastro}
\tablehead{\colhead{WD Name}&\colhead{R.A.}&\colhead{Dec.}&\colhead{Sep.}&\colhead{$\varpi$}&\colhead{$\mu_{\alpha}$}&\colhead{$\mu_{\delta}$}&\colhead{$\mu$ Ref.}&\colhead{\texttt{CoMover}}\\
&\colhead{($\degr$)}&\colhead{($\degr$)}&\colhead{($\arcsec$)}&\colhead{(mas)}&\colhead{(mas yr$^{-1}$)}&\colhead{(mas yr$^{-1}$)}&\colhead{}&\colhead{($\%$)}}
 \startdata
WD J001237.06$-$492422.81\tablenotemark{$\ddagger$}&3.15476907&-49.40629603&&10.4021$\pm$0.4282&52.587$\pm$0.256&9.574$\pm$0.373&1&\\
&3.15880278&-49.39958292&25.949&\dots&27.378$\pm$1.757&3.111$\pm$1.735&6& 0.0 \\ \hline
WD J001419.17$-$475719.58\tablenotemark{a}&3.58049053&-47.95620526&&11.0341$\pm$0.1946&94.839$\pm$0.083&-172.360$\pm$0.147&1&\\
&3.57858816&-47.95643084&4.656&11.7522$\pm$0.7092&94.336$\pm$0.315&-175.836$\pm$0.606&1& 100.0 \\ \hline
WD J014953.55$-$612916.97&27.47317349&-61.48876565&&13.0666$\pm$0.0544&3.697$\pm$0.060&-161.535$\pm$0.063&1&\\
&27.47055037&-61.48558188&12.316&13.1958$\pm$0.6313&5.231$\pm$0.721&-161.853$\pm$0.900&1& 100.0 \\ \hline
WD J023020.88$-$344749.38\tablenotemark{b}&37.58748692&-34.79751953&&13.5645$\pm$0.2254&88.835$\pm$0.140&-106.153$\pm$0.212&1&\\
&37.59010968&-34.79674285&8.242&13.8045$\pm$0.2157&88.687$\pm$0.126&-106.918$\pm$0.199&1& 100.0 \\ \hline
WD J023703.94$-$193253.64&39.26657798&-19.54812689&&12.8769$\pm$0.2640&35.157$\pm$0.262&24.646$\pm$0.269&1&\\
&39.26705588&-19.53837468&35.145&13.4481$\pm$0.1809&34.478$\pm$0.179&24.248$\pm$0.185&1& 100.0 \\ \hline
WD J032438.66$+$602055.88\tablenotemark{$\dagger$}&51.16129325&60.34885745&&14.905$\pm$0.0413&22.585$\pm$0.044&0.794$\pm$0.045&1&\\
&51.16544189&60.34854642&7.554&\dots&-14.040$\pm$8.160&2.900$\pm$8.160&4& 57.4 \\ \hline
WD J035637.43$+$321205.01\tablenotemark{$\dagger$}&59.15615247&32.20120185&&10.8855$\pm$0.1823&40.655$\pm$0.219&-42.978$\pm$0.160&1&\\
&59.15685548&32.20250415&5.078&\dots&-13.810$\pm$10.840&-25.720$\pm$10.840&4& 1.5 \\ \hline
WD J040046.84$-$261127.88\tablenotemark{bc}&60.19489510&-26.19155300&&10.1351$\pm$0.2233&-53.982$\pm$0.142&-105.948$\pm$0.237&1&\\
&60.19402800&-26.19183500&2.594&\dots&-55.980$\pm$6.250&-102.780$\pm$6.280&4& 100.0\\ \hline
WD J042641.33$+$404441.95&66.67226689&40.74489314&&11.2170$\pm$0.1776&8.511$\pm$0.206&-21.259$\pm$0.163&1&\\
&66.67143120&40.74522229&2.556&\dots&21.290$\pm$4.380&-23.970$\pm$4.280&4& 99.7 \\ \hline
WD J044713.58$-$582318.04\tablenotemark{d}&71.80782392&-58.38735132&&11.9753$\pm$0.0804&144.688$\pm$0.106&223.410$\pm$0.112&1&\\
&71.80068429&-58.38746020&13.478&11.8855$\pm$0.2510&143.481$\pm$0.324&222.571$\pm$0.356&1& 100.0 \\ \hline
WD J052555.93$-$132306.14&81.48317336&-13.38457114&&12.7147$\pm$0.1510&25.155$\pm$0.137&105.459$\pm$0.148&1&\\
&81.48324826&-13.38423666&1.232&12.8550$\pm$0.3505&20.137$\pm$0.439&103.773$\pm$0.379&1& 100.0\\ \hline
WD J065247.36$+$562851.61\tablenotemark{$\dagger$}&103.19740550&56.48103159&&10.1117$\pm$0.3520&10.907$\pm$0.245&6.823$\pm$0.200&1&\\
&103.19782560&56.48043620&2.460&\dots&9.260$\pm$7.360&29.000$\pm$7.360&2& 99.6 \\ \hline
WD J071609.53$+$003845.42&109.03985530&0.64555161&&10.0449$\pm$0.5858&28.578$\pm$0.543&-90.304$\pm$0.491&1&\\
&109.039150&0.644984&3.358&\dots&31.280$\pm$3.600&-100.750$\pm$3.500&2& 99.9 \\ \hline
WD J073504.07$-$794410.69\tablenotemark{$\dagger$}&113.76710360&-79.73633667&&10.3760$\pm$0.0434&5.641$\pm$0.058&-7.907$\pm$0.060&1&\\
&113.76953590&-79.73515280&4.640&\dots&37.750$\pm$46.100&23.610$\pm$42.700&3& 99.0 \\ \hline
WD J080648.05$+$221551.65\tablenotemark{be}&121.70014940&22.26354315&&20.2306$\pm$0.1295&-11.125$\pm$0.126&-181.042$\pm$0.080&1&\\
&121.70170060&22.26234460&7.787&\dots&-18.300$\pm$11.340&-183.470$\pm$12.060&1& 99.9 \\ \hline
WD J080700.06$-$535049.78&121.74985732&-53.84710492&&13.9069$\pm$0.2493&-50.375$\pm$0.268&11.883$\pm$0.317&1&\\
&121.75008980&-53.84741704&1.227&13.7964$\pm$1.0883&-44.250$\pm$1.195&15.130$\pm$1.491&1& 100.0 \\ \hline
WD J081319.85$-$135825.24\tablenotemark{d}&123.33260600&-13.97363171&&13.0609$\pm$0.1494&-26.206$\pm$0.151&10.691$\pm$0.157&1&\\
&123.33419840&-13.97350404&5.582&\dots&-35.280$\pm$4.290&9.440$\pm$4.290&4& 99.8 \\ \hline
WD J082459.63$-$254652.79\tablenotemark{b}&126.24805090&-25.78062645&&10.6518$\pm$0.3647&-86.185$\pm$0.282&157.933$\pm$0.262&1&\\
&126.24738718&-25.77977101&3.757&9.9408$\pm$0.6354&-84.208$\pm$0.450&157.595$\pm$0.444&1& 100.0 \\ \hline
WD J095004.78$+$130937.82&147.51965482&13.16008135&&17.7597$\pm$0.1563&-55.338$\pm$0.187&-95.202$\pm$0.225&1&\\
&147.51945820&13.16069700&2.078&\dots&-40.830$\pm$27.200&-126.450$\pm$29.000&3& 99.9 \\ \hline
WD J104627.91$+$605448.32&161.61635690&60.91362902&&21.6187$\pm$0.0972&8.503$\pm$0.078&46.585$\pm$0.088&1&\\
&161.61606317&60.91473344&4.009&21.5812$\pm$0.2066&5.318$\pm$0.157&43.842$\pm$0.193&1& 100.0 \\ \hline
WD J123405.10$+$442750.07\tablenotemark{$\dagger$}&188.52092740&44.46374053&&10.1323$\pm$0.3345&-48.873$\pm$0.198&-37.172$\pm$0.260&1&\\
&188.51936360&44.46728250&13.286&\dots&-29.450$\pm$8.330&-5.660$\pm$8.430&2& 14.8 \\ \hline
WD J125044.44$+$415259.50&192.68485650&41.88325499&&10.0430$\pm$0.8057&-52.491$\pm$0.629&11.303$\pm$0.662&1&\\
&192.6844836&41.8831214&1.229&\dots&-79.470$\pm$7.430&7.990$\pm$7.700&2& 98.2 \\ \hline
WD J134422.33$-$573824.60&206.09217325&-57.64030013&&11.3164$\pm$0.1938&-102.910$\pm$0.159&-30.165$\pm$0.180&1&\\
&206.08682414&-57.64129857&10.915&11.7255$\pm$0.4434&-103.885$\pm$0.378&-31.290$\pm$0.400&1& 100.0 \\ \hline
WD J141835.29$+$092919.68\tablenotemark{b}&214.64691690&9.48868440&&11.5405$\pm$0.0449&-29.125$\pm$0.056&-26.022$\pm$0.047&1&\\
&214.645675&9.489645&5.604&\dots&-29.420$\pm$4.980&-25.070$\pm$5.510&5& 100.0 \\ \hline
WD J164220.31$+$515150.98\tablenotemark{a}&250.58431770&51.86411101&&10.1741$\pm$0.3777&-39.625$\pm$0.478&-12.588$\pm$0.474&1&\\
&250.58220948&51.86437633&4.783&\dots&-40.910$\pm$3.970&-11.710$\pm$3.960&4& 100.0 \\ \hline
WD J171508.30$-$191029.62\tablenotemark{d}&258.78452710&-19.17520009&&15.7889$\pm$0.2672&-11.735$\pm$0.334&-68.499$\pm$0.226&1&\\
&258.78511871&-19.17524133&2.017&15.7888$\pm$0.4579&-10.815$\pm$0.582&-64.733$\pm$0.399&1& 100.0 \\ \hline
WD J200550.39$+$693227.22\tablenotemark{f}&301.45962820&69.54122246&&14.9830$\pm$0.1668&-25.056$\pm$0.234&74.422$\pm$0.205&1&\\
&301.45890459&69.54182851&2.364&14.6706$\pm$0.2874&-22.079$\pm$0.396&73.001$\pm$0.372&1& 100.0 \\ \hline
WD J202533.25$+$351510.15\tablenotemark{b}&306.38844874&35.25250972&&10.9730$\pm$0.0746&-18.947$\pm$0.071&-69.407$\pm$0.082&1&\\
&306.38957885&35.25067367&7.398&10.8499$\pm$0.4254&-18.576$\pm$0.373&-70.040$\pm$0.462&1& 100.0 \\ \hline
WD J211901.60$+$420617.34\tablenotemark{$\ddagger$}&319.75682370&42.10456425&&10.1232$\pm$0.0392&28.294$\pm$0.043&-56.703$\pm$0.041&1&\\
&319.74845991&42.09883801&30.398&11.6371$\pm$0.5564&21.572$\pm$0.599&-20.867$\pm$0.576&1& 0.0 \\ \hline
WD J235319.73$-$432517.86\tablenotemark{g}&358.33200230&-43.42287846&&14.1042$\pm$0.5174&-34.803$\pm$0.470&-283.899$\pm$0.433&1&\\
&358.33375000&-43.42238880&4.955&\dots&-36.970$\pm$16.680&-282.050$\pm$16.650&6& 100.0 \\ \hline
WD J235521.59$+$513106.10&358.84052670&51.51811023&&10.5125$\pm$0.3091&81.313$\pm$0.270&-56.934$\pm$0.272&1&\\
&358.84174019&51.51336131&17.311&10.7942$\pm$0.7163&82.521$\pm$0.609&-56.367$\pm$0.761&1& 100.0 \\
\enddata
\tablenotetext{\dagger}{Proper motions for these sources are not precise enough to firmly rule out or confirm physical association.}
\tablenotetext{\ddagger}{These systems are unlikely to be physical pairs.  The possible companion to WD J211901.60$+$420617.34 was followed-up spectroscopically (Section \ref{sec:obs}), despite its low \texttt{CoMover} probability.}
\tablenotetext{a}{Identified by Backyard Worlds citizen scientist L\'eopold Gramaize.}
\tablenotetext{b}{Identified by Backyard Worlds citizen scientist Frank Kiwy.}
\tablenotetext{c}{Identified by Backyard Worlds citizen scientist Tom Bickle.}
\tablenotetext{d}{Identified by Backyard Worlds citizen scientist Awab Abdullahi.}
\tablenotetext{e}{Identified by Backyard Worlds citizen scientist Christopher Tanner.}
\tablenotetext{f}{Identified by Backyard Worlds citizen scientist Arttu Sainio.}
\tablenotetext{g}{Identified by Backyard Worlds citizen scientist Austin Rothermich.}
\tablerefs{(1) Gaia DR3 \citep{gaia2021}; (2) UHS \citep{dye2018, schneider2023}; (3) CatWISE 2020 \citep{marocco2021}; (4) Pan-STARRS DR1 \citep{chambers2016}, This work; (5) UKIDSS LAS \citep{lawrence2007}, This work; (6) NSC DR2 \citep{nidever2021}}
\end{deluxetable*}
\end{longrotatetable}

\begin{longrotatetable}
\begin{deluxetable*}{cccccccccccccccccccccccccccccc}
\tabletypesize{\scriptsize}
\tablecaption{Optical Photometry for Resolved Companions}
\label{tab:resoptphoto}
\tablehead{\colhead{WD Name} & \colhead{$G$} & \colhead{$G_{\rm BP}$} & \colhead{$G_{\rm RP}$} & \colhead{$g$} & \colhead{$r$} & \colhead{$i$} & \colhead{$z$} & \colhead{$y$} & \colhead{Optical Source} & \\
 & \colhead{(mag)} & \colhead{(mag)} & \colhead{(mag)} & \colhead{(mag)} & \colhead{(mag)} & \colhead{(mag)} & \colhead{(mag)} & \colhead{(mag)} & \colhead{} &} 
\startdata
WD J0012$-$4924B & 21.1058 $\pm$ 0.0119 & 21.4001 $\pm$ 0.1770 & 19.7325 $\pm$ 0.1052 & 24.0977 $\pm$ 0.1465 & 22.3258 $\pm$ 0.0494 & 20.1014 $\pm$ 0.0130 & 19.1525 $\pm$ 0.0081 & 18.9283 $\pm$ 0.0173 & 1 & \\
WD J0014$-$4757B & 20.3254 $\pm$ 0.0050 & 21.7254 $\pm$ 0.1284 & 18.8303 $\pm$ 0.0226 & 21.2674 $\pm$ 0.0123 & 21.8619 $\pm$ 0.0287 & 19.1716 $\pm$ 0.0041 & 17.9558 $\pm$ 0.0026 & 17.5150 $\pm$ 0.0054 & 1 & \\
WD J0149$-$6129B & 20.5739 $\pm$ 0.0071 & 20.8254 $\pm$ 0.4157 & 19.1726 $\pm$ 0.0676 & 25.1062 $\pm$ 0.4038 & 22.2604 $\pm$ 0.0402 & 19.6117 $\pm$ 0.0064 & 18.1167 $\pm$ 0.0031 & 17.6817 $\pm$ 0.0062 & 1 & \\
WD J0230$-$3447B & 18.9590 $\pm$ 0.0033 & 21.3322 $\pm$ 0.0995 & 17.4195 $\pm$ 0.0088 & 22.3405 $\pm$ 0.0317 & 20.5743 $\pm$ 0.0083 & 17.9188 $\pm$ 0.0014 & 16.7030 $\pm$ 0.0009 & 16.3236 $\pm$ 0.0023 & 1 & \\
WD J0237$-$1932B & 18.7291 $\pm$ 0.0034 & 21.0786 $\pm$ 0.1415 & 17.2676 $\pm$ 0.0113 & 21.8223 $\pm$ 0.0836 & 20.5056 $\pm$ 0.0298 & 18.0492 $\pm$ 0.0068 & 16.9093 $\pm$ 0.0042 & 16.2883 $\pm$ 0.0045 & 2 & \\
WD J0324$+$6020B & \dots & \dots & \dots & \dots & \dots & 21.0822 $\pm$ 0.0690 & 19.6702 $\pm$ 0.0238 & 18.6915 $\pm$ 0.0230 & 2 & \\
WD J0356$+$3212B & \dots & \dots & \dots & \dots & \dots & 21.6415 $\pm$ 0.0727 & 20.5789 $\pm$ 0.0774 & 19.7736 $\pm$ 0.0351 & 2 & \\
WD J0400$-$2611B & \dots & \dots & \dots & 23.6926 $\pm$ 0.0885 & 22.4884 $\pm$ 0.0381 & 20.3585 $\pm$ 0.0108 & 18.8640 $\pm$ 0.0053 & 18.4100 $\pm$ 0.0118 & 1 & \\
WD J0426$+$4044B & \dots & \dots & \dots & $>$23.5590 & 21.3570 & 20.5831 $\pm$ 0.0890 & 19.1689 $\pm$ 0.0115 & 18.2433 $\pm$ 0.0273 & 2 & \\
WD J0447$-$5823B & 19.2811 $\pm$ 0.0044 & 21.0895 $\pm$ 0.1391 & 17.7795 $\pm$ 0.0142 & 22.9239 $\pm$ 0.0616 & 20.9196 $\pm$ 0.0125 & 18.2390 $\pm$ 0.0023 & 17.0908 $\pm$ 0.0014 & 16.7726 $\pm$ 0.0034 & 1 & \\
WD J0525$-$1323B & 19.2593 $\pm$ 0.0041 & \dots & \dots & 23.2540 $\pm$ 0.0784 & 20.8534 $\pm$ 0.0174 & 18.1790 $\pm$ 0.0016 & 17.0438 $\pm$ 0.0015 & \dots & 3 & \\    
WD J0652$+$5628B & \dots & \dots & \dots & $>$24.7740 & $>$22.3570 & $>$20.8190 & 19.7758 $\pm$ 0.0438 & 18.7978 $\pm$ 0.0234 & 2 & \\
WD J0716$+$0038B & \dots & \dots & \dots & $>$23.1870 & $>$21.1490 & 20.9548 $\pm$ 0.0415 & 19.4493 $\pm$ 0.0301 & 18.4756 $\pm$ 0.0270 & 2 & \\ 
WD J0735$-$7944B & \dots & \dots & \dots & 25.6991 $\pm$ 0.3062 & 23.8785 $\pm$ 0.0915 & 21.5762 $\pm$ 0.0172 & 20.5244 $\pm$ 0.0218 & \dots & 3 & \\
WD J0806$+$2215B & \dots & \dots & \dots & \dots & \dots & 21.5088 $\pm$ 0.0695 & 20.0639 $\pm$ 0.0172 & 18.9666 $\pm$ 0.0199 & 2 & \\
WD J0807$-$5350B & 20.6217 $\pm$ 0.0111 & \dots & \dots & \dots & \dots & \dots & 17.854 $\pm$ 0.046 & \dots & 4 & \\
WD J0813$-$1358B & \dots & \dots & \dots & $>$22.5200 & $>$20.7150 & 21.0905 $\pm$ 0.0592 & 19.7224 $\pm$ 0.0296 & 18.7183 $\pm$ 0.0210 & 2 & \\ 
WD J0824$-$2546B & 20.1222 $\pm$ 0.0066 & 20.8530 $\pm$ 0.1260 & 18.6103 $\pm$ 0.0302 & \dots & \dots & 19.5073 $\pm$ 0.0111 & 18.1973 $\pm$ 0.0217 & 17.5394 $\pm$ 0.0068 & 2 & \\
WD J0950$+$1309B & \dots & \dots & \dots & \dots & \dots & 23.9706 $\pm$ 0.3467 & 20.8270 $\pm$ 0.0288 & \dots & 3 & \\ 
WD J1046$+$6054B & 18.9159 $\pm$ 0.0034 & 21.3125 $\pm$ 0.1416 & 17.3298 $\pm$ 0.0091 & $>$20.1350 & 21.1886 $\pm$ 0.0560 & 18.3831 $\pm$ 0.0060 & 16.9785 $\pm$ 0.0069 & 16.1717 $\pm$ 0.0091 & 2 & \\
WD J1234$+$4427B & \dots & \dots & \dots & \dots & \dots & 20.4664 $\pm$ 0.0329 & 19.6619 $\pm$ 0.0203 & 19.1481 $\pm$ 0.0215 & 2 & \\  
WD J1250$+$4152B & \dots & \dots & \dots & \dots & 21.0860 $\pm$ 0.2588 & \dots & 20.7318 $\pm$ 0.0861 & 19.5414 $\pm$ 0.0450 & 2 & \\
WD J1344$-$5738B & 19.8849 $\pm$ 0.0045 & \dots & \dots & \dots & \dots & 18.8040 $\pm$ 0.0095 & 17.4591 $\pm$ 0.0056 & 16.8306 $\pm$ 0.0078 & 5 & \\
WD J1418$+$0929B & \dots & \dots & \dots & $>$24.4540 & $>$22.0270 & 20.9369 $\pm$ 0.0370 & 19.5623 $\pm$ 0.0250 & 18.6138 $\pm$ 0.0178 & 2 & \\
WD J1642$+$5151B & \dots & \dots & \dots & $>$21.7110 & $>$20.2500 & 20.8786 $\pm$ 0.0226 & 19.4321 $\pm$ 0.0127 & 18.4942 $\pm$ 0.0232 & 2 & \\ 
WD J1715$-$1910B & 19.8196 $\pm$ 0.0048 & 21.7778 $\pm$ 0.2158 & 18.1524 $\pm$ 0.0310 & $>$20.8400 & $>$19.1840 & 19.3805 $\pm$ 0.0242 & 17.8437 $\pm$ 0.0077 & 17.0125 $\pm$ 0.0093 & 2 & \\
WD J2005$+$6932B & 19.7230 $\pm$ 0.0046 & 21.0920 $\pm$ 0.3113 & 18.1748 $\pm$ 0.0298 & \dots & \dots & 19.1485 $\pm$ 0.0123 & 17.7141 $\pm$ 0.0099 & 16.9559 $\pm$ 0.0089 & 2 & \\ 
WD J2025$+$3515B & 20.0151 $\pm$ 0.0050 & 21.5649 $\pm$ 0.1600 & 18.4568 $\pm$ 0.0198 & $>$20.5270 & $>$18.9160 & 19.4778 $\pm$ 0.0239 & 18.0853 $\pm$ 0.0082 & 17.3116 $\pm$ 0.0101 & 2 & \\
WD J2119$+$4206B & 20.3323 $\pm$ 0.0064 & 21.6090 $\pm$ 0.3051 & 18.7856 $\pm$ 0.0219 & \dots & 21.0458 $\pm$ 0.0794 & 19.8073 $\pm$ 0.0140 & 18.3919 $\pm$ 0.0089 & 17.5155 $\pm$ 0.0105 & 2 & \\
WD J2353$-$4325B & \dots & \dots & \dots & 27.5550 $\pm$ 4.6653 & 24.3429 $\pm$ 0.3295 & 24.0255 $\pm$ 0.4225 & 21.5718 $\pm$ 0.0873 & 20.9234 $\pm$ 0.1308 & 1 & \\
WD J2355$+$5131B & 20.4489 $\pm$ 0.0059 & 21.2955 $\pm$ 0.1300 & 18.8772 $\pm$ 0.0255 & \dots & \dots & 19.9599 $\pm$ 0.0332 & 18.5047 $\pm$ 0.0144 & 17.7087 $\pm$ 0.0171 & 2 & \\ 
\enddata
\tablerefs{(1) The Dark Energy Survey (DES; \citealt{des2005}); (2) PS1 \citep{chambers2016}; (3) DESI Legacy Imaging Surveys \citep{dey2019}; (4) SkyMapper \citep{wolf2018}; (5) NSC DR2 \citep{nidever2021}}
\end{deluxetable*}
\end{longrotatetable}

\begin{deluxetable*}{cccccccccccccccccccccccccccccc}
\tabletypesize{\scriptsize}
\tablecaption{Infrared Photometry for Resolved Companions}
\label{tab:resirphoto}
\tablehead{
\colhead{WD Name} & \colhead{$J$} & \colhead{$H$} & \colhead{$K$} & \colhead{NIR Source} & \colhead{W1} & \colhead{W2} \\
& \colhead{(mag)} & \colhead{(mag)} & \colhead{(mag)} & \colhead{} & \colhead{(mag)} & \colhead{(mag)}}
\startdata
WD J0012$-$4924B&17.419$\pm$0.025&16.987$\pm$0.044&16.592$\pm$0.053&1&16.437$\pm$0.032&16.269$\pm$0.074\\
WD J0014$-$4757B&15.790$\pm$0.006&15.210$\pm$0.006&14.709$\pm$0.008&1&14.320$\pm$0.014&14.204$\pm$0.015\\
WD J0149$-$6129B&15.938$\pm$0.007&\dots&15.126$\pm$0.013&1&14.746$\pm$0.014&14.391$\pm$0.015\\
WD J0230$-$3447B&14.951$\pm$0.040&14.355$\pm$0.044&13.913$\pm$0.051&2&13.740$\pm$0.015&13.513$\pm$0.012\\
WD J0237$-$1932B&14.748$\pm$0.004&\dots&13.877$\pm$0.006&1&13.685$\pm$0.015&13.476$\pm$0.012\\
WD J0324$+$6020B&16.626$\pm$0.126&15.841$\pm$0.155&15.091$\pm$0.117&2&17.528$\pm$0.086&17.784$\pm$0.297\\
WD J0356$+$3212B&18.367$\pm$0.074&\dots&17.398$\pm$0.064&3&\dots&\dots\\
WD J0400$-$2611B&16.543$\pm$0.008&\dots&15.517$\pm$0.018&1&15.136$\pm$0.018&14.977$\pm$0.026\\
WD J0426$+$4044B&\dots&\dots&15.285$\pm$0.009&3&14.898$\pm$0.017&14.821$\pm$0.027\\
WD J0447$-$5823B&15.258$\pm$0.004&\dots&14.384$\pm$0.007&1&14.177$\pm$0.012&13.960$\pm$0.011\\
WD J0525$-$1323B&15.167$\pm$0.004&\dots&14.350$\pm$0.008&1&14.066$\pm$0.015&13.876$\pm$0.015\\
WD J0652$+$5628B&16.834$\pm$0.021&\dots&15.614$\pm$0.028&4&15.211$\pm$0.022&15.046$\pm$0.028\\
WD J0716$+$0038B&16.540$\pm$0.016&\dots&15.440$\pm$0.022&4&15.125$\pm$0.018&15.128$\pm$0.033\\
WD J0735$-$7944B&18.568$\pm$0.092&\dots&17.452$\pm$0.159&1&17.083$\pm$0.052&16.985$\pm$0.123\\
WD J0806$+$2215B&16.988$\pm$0.153&15.847$\pm$0.131&14.901$\pm$0.111&2&14.406$\pm$0.017&14.227$\pm$0.019\\
WD J0807$-$5350B&15.902$\pm$0.010&\dots&15.024$\pm$0.017&1&14.630$\pm$0.014&14.371$\pm$0.015\\
WD J0813$-$1358B&16.615$\pm$0.010&\dots&15.486$\pm$0.079&1&15.056$\pm$0.018&14.924$\pm$0.027\\
WD J0824$-$2546B&15.959$\pm$0.008&\dots&15.080$\pm$0.016&1&14.723$\pm$0.021&14.535$\pm$0.023\\
WD J0950$+$1309B&17.871$\pm$0.033&17.265$\pm$0.043&16.989$\pm$0.064&3& \dots & \dots \\
WD J1046$+$6054B&14.534$\pm$0.030&13.876$\pm$0.041&13.358$\pm$0.033&2&13.041$\pm$0.011&12.825$\pm$0.009\\
WD J1234$+$4427B&17.788$\pm$0.044&\dots&17.020$\pm$0.067&4&16.721$\pm$0.035&16.423$\pm$0.077\\
WD J1250$+$4152B&17.592$\pm$0.037&\dots&16.623$\pm$0.062&4&16.056$\pm$0.025&15.825$\pm$0.043\\
WD J1344$-$5738B&15.611$\pm$0.083&14.993$\pm$0.095&14.924$\pm$0.141&2&14.272$\pm$0.013&14.260$\pm$0.015\\
WD J1418$+$0929B&16.613$\pm$0.010&15.905$\pm$0.011&15.313$\pm$0.012&3&15.092$\pm$0.026&14.935$\pm$0.034\\
WD J1642$+$5151B&16.753$\pm$0.127&16.178$\pm$0.201&\dots&2&15.218$\pm$0.015&14.944$\pm$0.019\\
WD J1715$-$1910B&15.262$\pm$0.005&\dots&14.286$\pm$0.008&1&13.872$\pm$0.018&13.655$\pm$0.016\\
WD J2005$+$6932B&\dots&\dots&\dots&\dots&14.102$\pm$0.013&13.828$\pm$0.011\\
WD J2025$+$3515B&15.600$\pm$0.005&\dots&\dots&3&14.041$\pm$0.043&13.826$\pm$0.043\\
WD J2119$+$4206B&15.826$\pm$0.005&\dots&14.668$\pm$0.007&3&14.555$\pm$0.018&14.413$\pm$0.022\\
WD J2353$-$4325B&18.901$\pm$0.050&\dots&18.738$\pm$0.247&1&17.206$\pm$0.056&16.177$\pm$0.063\\
WD J2355$+$5131B&16.021$\pm$0.010&\dots&14.967$\pm$0.013&4&14.751$\pm$0.017&14.546$\pm$0.019\\
\enddata
\tablerefs{(1) VHS \citep{mcmahon2013}; (2) 2MASS \citep{skrutskie2006}; (3) UKIDSS \citep{lawrence2007}; (4) UHS \citep{dye2018}}
\end{deluxetable*}

\begin{deluxetable*}{ccccc}
\tabletypesize{\footnotesize}
\tablecaption{Candidate Companion Spectral Types}
\label{tab:sptest}
\tablehead{\colhead{WD Name} & \colhead{$M_G$} & \colhead{$M_J$} & \colhead{$M_K$} & \colhead{Spectral\tablenotemark{a}}\\
&\colhead{(mag)} & \colhead{(mag)} & \colhead{(mag)} & \colhead{Type}}
 \startdata
WD J0012$-$4924B&16.191$\pm$0.090&12.505$\pm$0.093&11.678$\pm$0.104&[L3]\\
WD J0014$-$4757B&15.676$\pm$0.131&11.141$\pm$0.131&10.060$\pm$0.131&[M8]\\
WD J0149$-$6129B&16.176$\pm$0.104&11.540$\pm$0.104&10.728$\pm$0.105&[L0]\\
WD J0230$-$3447B&14.659$\pm$0.034&10.651$\pm$0.052&9.613$\pm$0.061&[M7]\\
WD J0237$-$1932B&14.371$\pm$0.029&10.391$\pm$0.029&9.520$\pm$0.030&[M7]\\
WD J0324$+$6020B&\dots&12.493$\pm$0.126&10.958$\pm$0.117&[L2]\\
WD J0356$+$3212B&\dots&13.551$\pm$0.082&12.582$\pm$0.074&[L6]\\
WD J0400$-$2611B&\dots&11.572$\pm$0.048&10.546$\pm$0.051&[L0]\\
WD J0426$+$4044B&\dots&\dots&10.534$\pm$0.036&[L0]\\
WD J0447$-$5823B&14.656$\pm$0.046&10.633$\pm$0.046&9.759$\pm$0.046&[M7]\\
WD J0525$-$1323B&14.805$\pm$0.059&10.712$\pm$0.059&9.895$\pm$0.060&[M7]\\
WD J0652$+$5628B&\dots&11.858$\pm$0.078&10.638$\pm$0.081&[L1]\\
WD J0716$+$0038B&\dots&11.550$\pm$0.128&10.450$\pm$0.128&[L0]\\
WD J0735$-$7944B&\dots&13.648$\pm$0.092&12.532$\pm$0.159&[L6]\\
WD J0807$-$5350B&16.321$\pm$0.172&11.601$\pm$0.171&10.723$\pm$0.172&[L0]\\
WD J0813$-$1358B&\dots&12.195$\pm$0.027&11.066$\pm$0.083&[L1]\\
WD J0824$-$2546B&15.109$\pm$0.139&10.946$\pm$0.139&10.067$\pm$0.140&[M8]\\
WD J0950$+$1309B&\dots&14.118$\pm$0.038&13.236$\pm$0.067&[L8]\\
WD J1234$+$4427B&\dots&12.817$\pm$0.084&12.049$\pm$0.098&[L3]\\
WD J1250$+$4152B&\dots&12.601$\pm$0.178&11.632$\pm$0.185&[L3]\\
WD J1344$-$5738B&15.231$\pm$0.082&10.957$\pm$0.117&10.270$\pm$0.163&[M8]\\
WD J1418$+$0929B&\dots&11.924$\pm$0.013&10.624$\pm$0.015&[L1]\\
WD J1642$+$5151B&\dots&11.790$\pm$0.150&\dots&[L0]\\
WD J1715$-$1910B&15.811$\pm$0.063&11.254$\pm$0.063&10.278$\pm$0.063&[M9]\\
WD J2005$+$6932B&15.555$\pm$0.043&\dots&\dots&[M9]\\
WD J2353$-$4325B&\dots&14.648$\pm$0.094&\dots&[T5]\\ \hline
WD J0806$+$2215B&\dots&13.518$\pm$0.154&11.431$\pm$0.112&L3\\
WD J1046$+$6054B&15.586$\pm$0.021&11.204$\pm$0.037&10.028$\pm$0.039&M9\tablenotemark{b}\\
WD J2025$+$3515B&15.192$\pm$0.085&10.777$\pm$0.085&\dots&M8\\
WD J2119$+$4206B&15.662$\pm$0.104&11.155$\pm$0.104&9.997$\pm$0.104&M8\\
WD J2355$+$5131B&15.615$\pm$0.144&11.187$\pm$0.144&10.133$\pm$0.145&M8\\ \hline
WD J1125$-$1631B&\dots&\dots&\dots&M7\tablenotemark{c}\\
WD J1657$-$1256B&\dots&\dots&\dots&M4\tablenotemark{c}\\
WD J2339$+$2552B&\dots&\dots&\dots&M7\tablenotemark{c}\\
\enddata
\tablenotetext{a}{Spectral types in square brackets are estimates based on the absolute magnitudes listed and spectral type versus absolute magnitude relations from \cite{pecaut2013}, \cite{kiman2019}, and \cite{schneider2023}, while those spectral types without square brackets are those that have been confirmed spectroscopically.}
\tablenotetext{b}{WD J1046$+$6054B has an optical spectral type of M9 from \cite{kiman2019}.}
\tablenotetext{c}{These spectral types were determined via spectral decomposition in Section \ref{sec:specbin}.}
\end{deluxetable*}

\section{Observations}
\label{sec:obs}

\subsection{IRTF/SpeX}
\begin{figure*}
\plotone{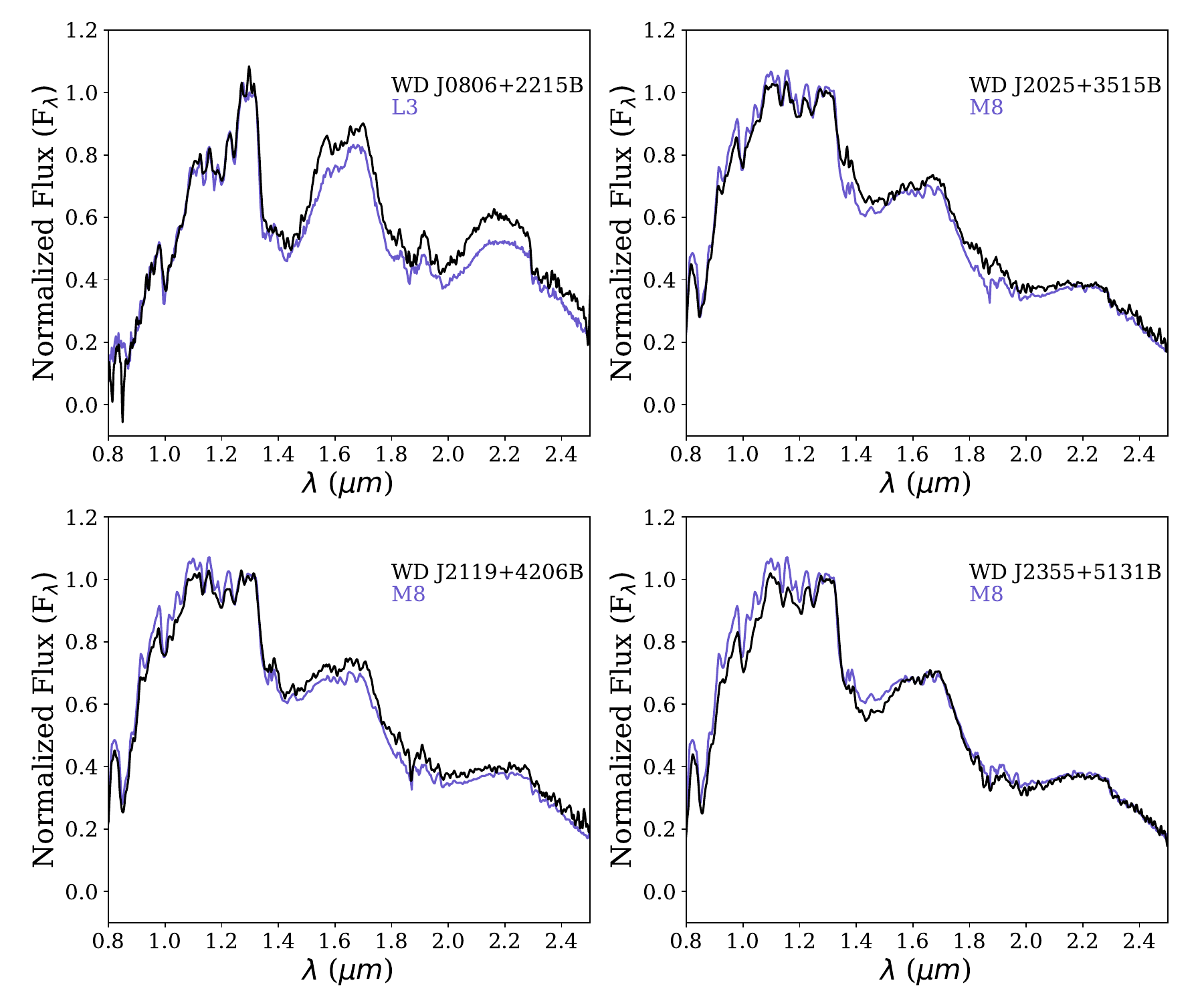}
\caption{IRTF/SpeX spectra of four newly discovered companions. All spectra are normalized between 1.27 and 1.29 $\mu$m.  WD J0806$+$2215B is best matched to the the L3 spectral standard (2MASSW J1506544+132106; \citealt{burgasser2007}), while WD J2025$+$3515B, WD J2119+4206B, and WD J2355$+$5131B are best matched to the M8 spectral standard (VB 10; \citealt{burgasser2004}). } 
\label{fig:spec1}
\end{figure*}

Four of our resolved candidate companions were observed with the SpeX spectrograph \citep{rayner2003} at NASA's Infrared Telescope Facility (IRTF). Observations for WD J0806$+$2215B were taken on 11 November 2018 and for WD J2025$+$3515B, WD J2119+4206B, and WD J2355$+$5131B on 13 August 2024. All observations were done in prism mode with the 0\farcs8 slit, which results in a resolving power of $\approx$200 across 0.8--2.5 $\mu$m. The spectra were reduced using the Spextool reduction package \citep{cushing2004, vacca2003}. The final reduced spectra is shown in Figure \ref{fig:spec1}. Comparing to near-infrared L and T dwarf spectral standards from \cite{burgasser2006} and \cite{kirkpatrick2010}, we find that the best match at the $J$-band for WD J0806$+$2215B is the L3 standard (2MASSW J1506544+132106; \citealt{burgasser2007}). For WD J2025$+$3515B, WD J2119+4206B, and WD J2355$+$5131B, we find the best matching standard at $J$-band to be the M8 standard (VB 10; \citealt{burgasser2004}).


\subsection{Gemini/GMOS}

\begin{figure}
\plotone{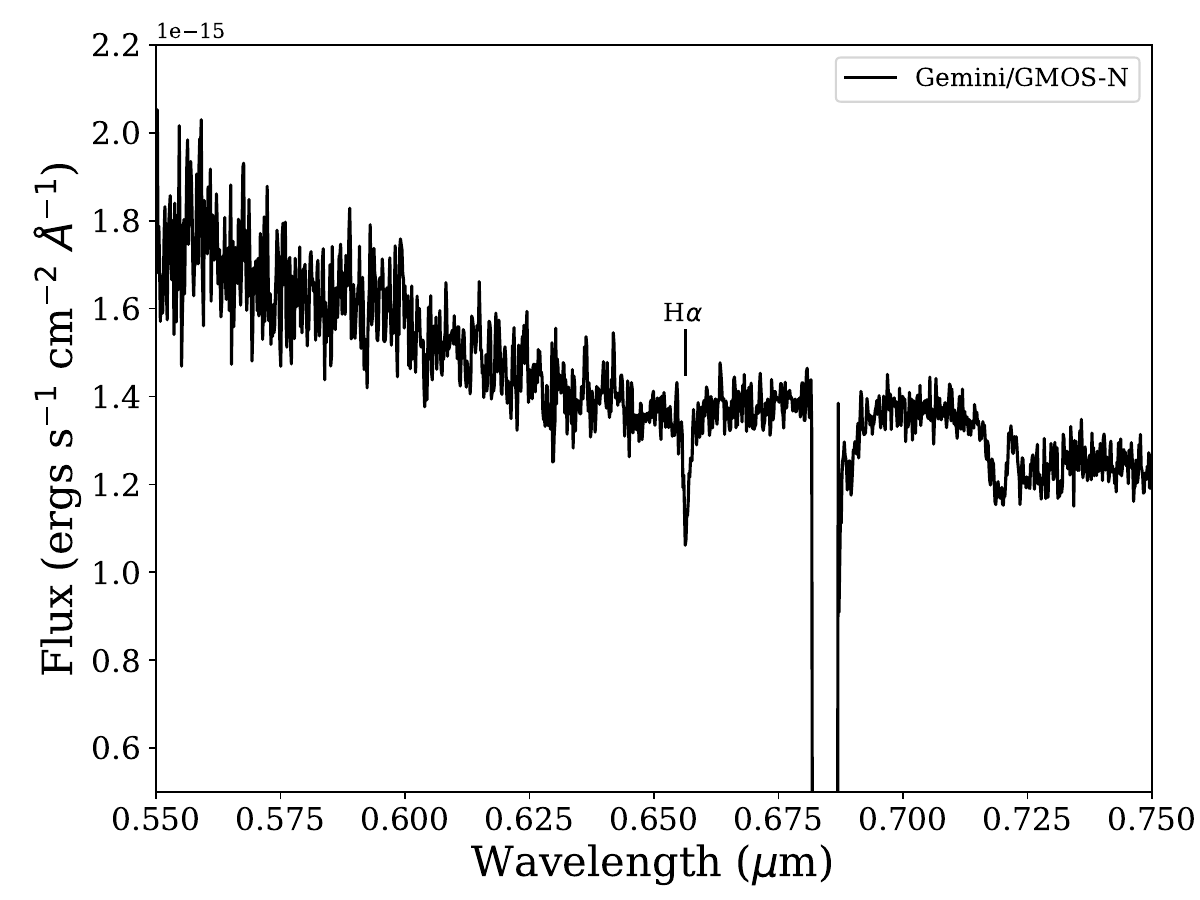}
\caption{Gemini/GMOS-N spectrum of WD J0400$-$2611 showing clear H$\alpha$ absorption and a lack of He lines.} 
\label{fig:wd0400}
\end{figure}

One white dwarf primary (WD J0400$-$2611) was observed with the Gemini-North Multi-Object Spectrograph (GMOS; \citealt{hook2004}) as part of program GN-2021A-Q-316 (PI: Debes). WD J0400$-$2611 was observed with the R400 grating using the 1\arcsec\ slit with 2$\times$2 binning centered on  0.764 $\mu$m.  The purpose of this observation was to examine the spectrum for the presence of hydrogen or helium to determine whether this object had a DA or non-DA composition.  We obtained 4, 300~s exposures that were then stacked during the reduction. The spectra were reduced using the \textsc{dragons} reduction software \citep{labrie2019, labrie2023}, which performs the bias, flat-field, bad pixel and wavelength corrections as well as the response function determined from a standard star. The H$\alpha$ absorption line, which indicates that this object is a DA white dwarf, is shown in Figure \ref{fig:wd0400}.

\section{System Properties}
\label{sec:sys}

\subsection{\teff\ and log(g)}
We gathered the basic fundamental properties of all candidate systems by searching the literature for \teff\ and log(g) values of the primary white dwarf and taking the most precise values available if more than one estimate existed.  Most of these measurements come from newly available low-resolution Gaia XP spectra (e.g., \citealt{jimenez2023, vincent2024}). These values are provided in Table \ref{tab:ages}.  

\subsection{Spectral Types}
We searched the literature for spectral types for the candidate unresolved systems and the white dwarf primaries of resolved co-moving systems. Spectral types, including our type found for WD J0400$-$2611 from Section \ref{sec:obs}, are given in Table \ref{tab:ages}. 

\subsection{Ages}
We determined age estimates for all of our candidate systems using \texttt{wdwarfdate} \citep{kiman2022}. \texttt{wdwarfdate} estimates the total ages of white dwarfs by combining their cooling ages with the main-sequence lifetimes of their progenitor stars. It uses the Initial-to-Final Mass Relation (IFMR), and in this study, we used the highest-accuracy \cite{cummings2018} (both MIST and PARSEC-based), which links the white dwarf's mass to the mass of its progenitor. The package incorporates cooling models from \cite{bedard2020} which detail the cooling process of white dwarfs based on their mass, temperature, and atmospheric composition. The package also incorporates stellar evolution models from \cite{dotter2016} and \cite{choi2016} to estimate the main-sequence lifetime of the progenitor star. \texttt{wdwarfdate} also handles DA and non-DA white dwarfs differently. For DA white dwarfs, the models from \cite{bedard2020} address their slower cooling rates, considering that the hydrogen layer is more insulating than a non-DA helium counterpart \citep{fontaine2001}. These non-DA counterparts also have their own distinct cooling model, so as to best estimate an age with models most representative of the actual composition of the object.   If no spectral type was found for an object in our sample, we evaluate both DA and non-DA ages for that system. Some objects had conflicting spectral types in the literature (e.g., WD J001419.17$-$475719.58 is typed as non-DA in \citealt{jimenez2023} and DA in \citealt{vincent2024}). In these cases, we evaluate both DA and non-DA scenarios and provide both age estimates.

Ages or age limits were determined for all but four of the unresolved candidates (Table \ref{tab:ages}). WD J1628$-$0418, WD J1657$-$1256, and WD J2053$+$2603 had no available \teff\ or log(g) values available in the literature, and WD J2339$+$2552 has an unusually low log(g) value (6.88$\pm$0.02; \citealt{vincent2024}) that is beyond the allowed values of the models considered by \texttt{wdwarfdate} \citep{kiman2022}.  Ages or age limits are estimated for all thirty-one resolved candidate systems (Table \ref{tab:ages}). 



\section{Analysis}
\label{sec:analysis}

\subsection{Spectral Binaries}
\label{sec:specbin}
\begin{figure*}
\plotone{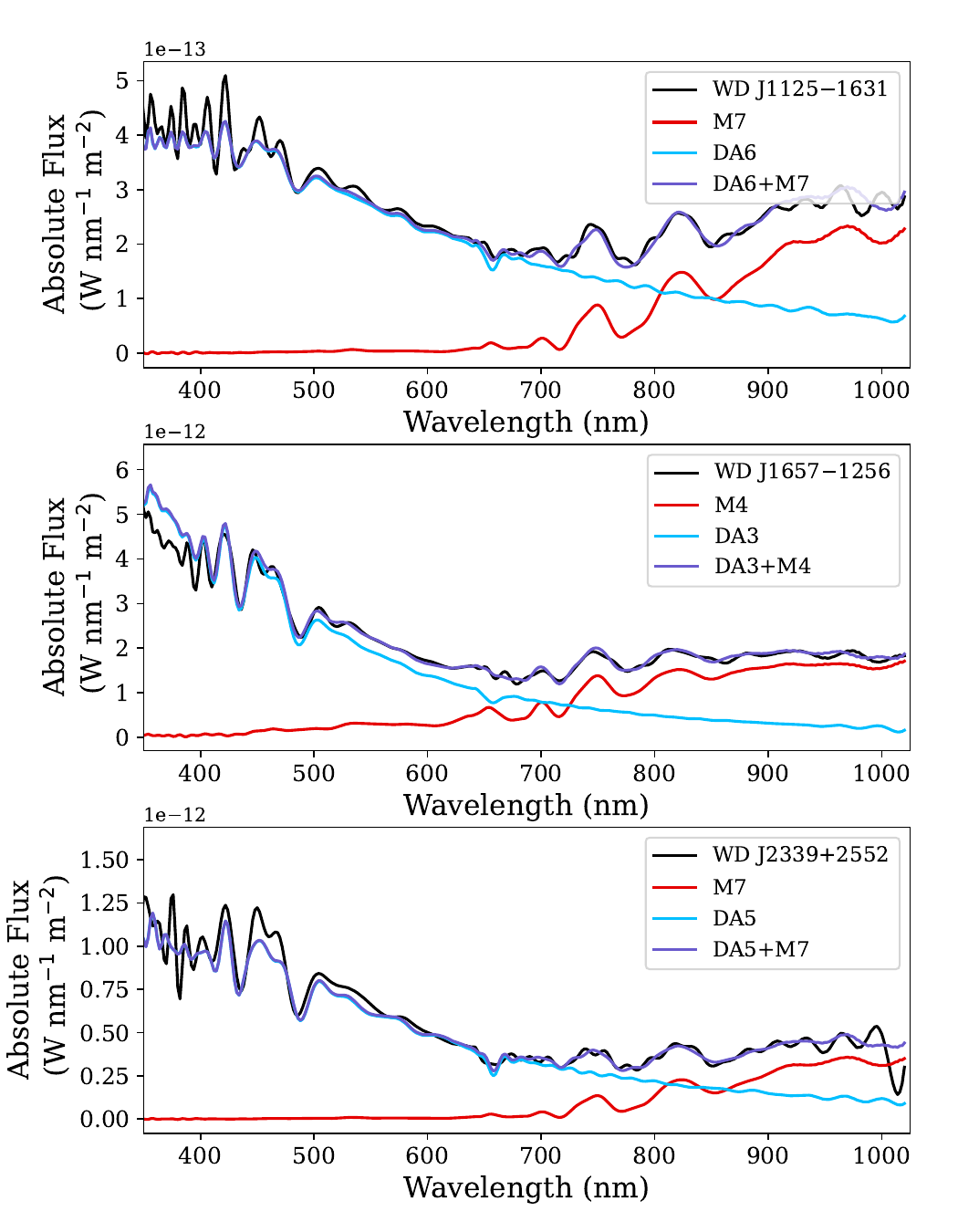}
\caption{Spectral binary fits to the Gaia spectra of WD J1125$-$1631, WD J1657$-$1256, and WD J2339$+$2552. White dwarf spectra are shown in cyan, M dwarf spectra are shown in red, and the combined white dwarf + M dwarf templates are shown in purple. The observed Gaia spectra for these three unresolved candidates are shown in black.} 
\label{fig:unrespec}
\end{figure*}

Several objects (WD J1125$-$1631AB, WD J1657$-$1256AB, and WD J2339$+$2552AB) have available optical spectra from Gaia DR3 \citep{gaia2021, deangeli2023}. We extracted each available spectrum using the GaiaXPy Python library \citep{ruz2024}. We inspected each spectrum for potential spectral signatures of low-mass companions. Three objects (WD J1125$-$1631, WD J1657$-$1256, and WD J2339$+$2552) showed clear signs of unresolved binarity with a low-mass companion. We made templates using spectroscopically classified white dwarfs from \cite{mccook1999} and M and L dwarf standards \citep{kirkpatrick1991, kirkpatrick1999}. All of the spectra were absolutely flux-calibrated using Gaia DR3 parallaxes, however we allow for additional multiplicative coeffecients to the white dwarf, M dwarf, and L dwarf templates to allow for unaccounted flux discrepancies such as variability or additional unresolved binarity of either the standard or the candidates. The best fitting templates were found via $\chi$$^2$-fitting to the white dwarf + cool dwarf templates. The white dwarfs used in the final fits are GD 140 (DA3; \citealt{mccook1999}), EGGR 155 (DA5; \citealt{mccook1999}), and EGGR 7 (DA6; \citealt{mccook1999}), while the M dwarfs are Gl 402 (M4; \citealt{kirkpatrick1991}) and VB 8 (M7; \citealt{henry2004}). The best fitting templates to the Gaia spectra of WD J1125$-$1631, WD J1657$-$1256, and WD J2339$+$2552 are shown in Figure \ref{fig:unrespec}.

\subsection{Old Systems}
WD J0012$-$4924AB, WD J0824$-$2546AB, WD J1046$+$6054AB, WD J1715$-$1910AB, WD J2005$+$6932AB, and WD J2353$-$4325AB all have age estimates $>$10 Gyr, making them potential benchmark systems with advanced ages. 
 
WD J1046$+$6054B has a confirmed optical spectral type of M9 from \cite{kiman2019}.  While no suggestion of low-metallicity was made for this source in that work, the calculated tangential velocity (\vtan) of $\sim$250 \kms\ supports the old age estimate for this system.  WD J0824$-$2546B, WD J1715$-$1910B, WD J2005$+$6932B all have estimated spectral types of late-M similar to WD J1046$+$6054B.  The WD J1715$-$1910AB system may be particularly valuable with the oldest age estimate in the entire sample presented in this work (11.83$^{+2.10}_{-2.08}$ Gyr).    

WD J0012$-$4924AB is a system that needs future astrometric confirmation of the physical association of the potential companion and the white dwarf.  If confirmed, the candidate companion has an estimated spectral type of $\sim$L3.  Follow-up spectroscopy could reveal this as a potential benchmark system containing an L subdwarf with a well-constrained age, joining a relatively rare group that includes VVV 1256$-$62AB \citep{zhang2024}.  

WD J2353$-$4325A does not have a confirmed white dwarf spectral type, but has an age of 11.12$^{+2.03}_{-1.92}$ Gyr based on the DA cooling model and 9.92$^{+2.90}_{-1.27}$ Gyr based on the non-DA model.  The companion in this system, WD J2353$-$4325B, has a spectral type estimate of $\sim$T5.  Therefore this system could be especially valuable as it contains a mid-T dwarf with an exceptionally old age.  The cloudless Sonora models \citep{marley2021} give a mass estimate of 70.4$\pm$3.3 \mjup\ and 69.9$\pm$2.8 \mjup\ for a T5 dwarf at the DA and non-DA age of the primary, respectively.

\subsection{Notes on Individual Systems}

\subsubsection{WD J0004$+$7401} 
Confirmation of infrared excess with higher resolution follow-up infrared imaging (e.g., \citealt{lai2021}) or additional photometric detections at longer wavelengths for this unresolved source can help to distinguish between true infrared excess due to a dusty debris disk or some contaminating flux from another source. None of our seventeen excess systems have additional infrared imaging from the Spitzer Space Telescope in the Spitzer Heritage Archive \citep{wu2010}. We further looked for longer wavelength detections at 12 $\mu$m and 22 $\mu$m in the AllWISE catalog \citep{kirkpatrick2014}, as the CatWISE2020 catalog only includes W1 and W2 photometry \citep{marocco2021}. Only three objects have 12 $\mu$m (W3) detections (WD J0004$+$7401, WD J1125$-$1631, and WD J1657$-$1256), and there are no detections at 22 $\mu$m (W4). The W3 detections of WD J1125$-$1631 and WD J1657$-$1256 are most likely due to unresolved stellar companions, as discussed in the next paragraph. The W3 detection of WD J0004$+$7401 may be evidence of circumstellar material for this source. Its extra-red nature is visible in Figure \ref{fig:cutouts2a}, with a reddish hue in the WISE composite image that is not seen for the other sources in that figure  due to the brighter than typical W3 flux for this source. Confirmation is possible with future high-resolution infrared images or mid- to far-infrared spectroscopy.  The total age for this white dwarf is $>$5.64$^{+4.57}_{-2.38}$ Gyr if a hydrogen dominated atmosphere and $>$6.08$^{+2.05}_{-2.69}$ Gyr in the non-DA case.  With a \teff\ $\sim$4300-4400 K, this would be one of the coldest and oldest white dwarfs to show infrared excess due to a dusty disk (e.g., \citealt{debes2019}).    

\subsubsection{WD J0024$-$1631}
This unresolved source is almost certainly a white dwarf + low-mass companion system, as a very red point source can be seen blended with the white dwarf primary in the PS1 images in Figure \ref{fig:cutouts2a}. However, this source was not resolved into two detections in the PS1 DR2 catalog at this position. Because this pair was not resolved in any existing catalog, it is included in our list of unresolved candidates. High-resolution imaging of this system and any resulting unblended photometry would allow for an accurate spectral type estimate for this companion.

\subsubsection{WD J0735$-$7944AB}
This white dwarf in this resolved system has the highest \teff\ in our sample (28063$\pm$836 K; \citealt{jimenez2023}) and consequently has the youngest age estimate (0.21$\pm$0.02 Gyr).  The high \teff\ of this source is supported by the values reported in \cite{gentile2021} (28583$\pm$1173 K) and \cite{vincent2024} (35337$\pm$1290 K).  We note that this is one of our ``unreliable'' systems for which a detailed astrometric analysis of the potential companion is needed.  However, if confirmed, we estimate a spectral type of $\sim$L6 for the putative companion.  At the age of WD J0735$-$7944A, an L6 companion would have a mass of 34$\pm$5 \mjup\ according to the solar metallicity hybrid-grav Sonora Diamondback models from \cite{morley2024}.  

\subsubsection{WD J0806$+$2215AB}
The companion in this resolved system was observed with IRTF/SpeX (Figure \ref{fig:spec1}) and was found to have a spectral type of L3.  The spectrum is slightly redder than the L3 standard, which is unlikely to be related to a young age (i.e., low surface gravity), as the age of the system is found to be 5.56$^{+3.08}_{-0.98}$ Gyr.  Other factors can cause redder colors, such as enhanced metallicity (e.g., \citealt{looper2008, zhang2024b}) or inclination angle (e.g., \citealt{vos2017, suarez2023}).  We find a \teff\ of 1841$\pm$147 K using the spectral type vs.\ \teff\ relations from \cite{kirkpatrick2021}, leading to a mass estimate of 77.4$\pm$2.6 \mjup\ using the age of the system and the solar-metallicity hybrid-grav models of \cite{morley2024}.  This is just above the hydrogen burning limit for the hybrid models (70.2 \mjup; \citealt{morley2024}) making this a likely low-mass, stellar companion. 

\subsubsection{WD J0950$+$1309AB}
Of the remaining resolved candidates with reliable astrometry that give a high probability of physical association with a white dwarf primary not already discussed previously, WD J0950$+$1309B is the lone object with a spectral type estimate clearly in the substellar regime ($\sim$L8).  Spectroscopic confirmation for the primary and secondary are needed for this system, which could become a benchmark object with a well-determined age at the L/T transition.

\subsubsection{WD J2119$+$4206}
The possible companion in this resolved system is all but ruled-out as physically associated by the \texttt{CoMover} results in Table \ref{tab:resastro}.  The companion candidate was followed-up spectroscopically (Section \ref{sec:obs}) before being evaluated with \texttt{CoMover}, and found to have a spectral type of M8.  This system is likely a chance alignment. 

\startlongtable
\begin{deluxetable*}{cccccccccccccccccccccccccccccc}
\tabletypesize{\footnotesize}
\tablecaption{Age Estimates for All Candidates}
\label{tab:ages}
\tablehead{\colhead{}&\colhead{Spectral}&\colhead{Ref.}&\colhead{$T_{\rm eff}$}&\colhead{$log(g)$}&\colhead{Ref.}&\colhead{Age}\\
\colhead{}&\colhead{Type}&\colhead{}&\colhead{(K)}&\colhead{(cgs)}&\colhead{}&\colhead{(Gyr)}} 
\startdata
WD J0004$+$7401 & {\it [DA]} & \dots & 4289$\pm$426 & 7.82$\pm$0.39 & 8 & $>$5.64$^{+4.57}_{-2.38}$ \\ 
 & {\it [non-DA]} & \dots & 4392$\pm$307 & 7.85$\pm$0.36 & 8 & $>$6.08$^{+2.05}_{-2.69}$ \\ \hline
WD J0012$-$4924 & DA & 1 & 4720$\pm$111 & 8.05$\pm$0.07 & 1 & 10.82$^{+2.42}_{-1.26}$ & \\ \hline
WD J0014$-$4757 & DA & 1 & 6420$\pm$93 & 8.24$\pm$0.03 & 1 & 3.93$^{+0.29}_{-0.27}$ \\ 
 & non-DA & 2 & 6035$\pm$125 & 8.15$\pm$0.05 & 2 & 5.26$^{+2.25}_{-0.71}$ \\ \hline
WD J0024$-$1631 & DA & 1 & 4608$\pm$150 & 7.77$\pm$0.09 & 1 & $>$4.18$^{+1.03}_{-0.72}$ \\ \hline
WD J0045$+$1454 & DA & 3 & 6963$\pm$77 & 8.38$\pm$0.02 & 1 & 3.80$^{+0.15}_{-0.16}$ \\ \hline
WD J0127$-$5759 & DA & 2 & 15888$\pm$426 & 8.95$\pm$0.02 & 1 & 1.07$^{+0.09}_{-0.08}$ \\ \hline
WD J0149$-$6129 & DQ & 1 & 8581$\pm$72 & 7.89$\pm$0.02 & 1 & $>$0.84$^{+0.02}_{-0.03}$ \\ \hline
WD J0230$-$3447 & DA & 1 & 4760$\pm$92 & 7.84$\pm$0.06 & 1 & $>$4.26$^{+0.88}_{-0.58}$ \\ \hline
WD J0237$-$1932 & DA & 1 & 5079$\pm$127 & 8.01$\pm$0.07 & 1 & 9.84$^{+3.08}_{-2.00}$ \\ \hline
WD J0324$+$6020 & DA & 1 & 10960$\pm$76 & 8.03$\pm$0.01 & 1 & 3.67$^{+3.94}_{-1.55}$ & \\ \hline
WD J0356$+$3212 & DA & 1 & 8040$\pm$223 & 8.56$\pm$0.04 & 1 & 3.32$^{+0.27}_{-0.26}$ & \\ \hline
WD J0400$-$2611 & DA & 11 & 5688$\pm$89 & 8.00$\pm$0.04 & 1 & 7.58$^{+4.06}_{-2.41}$ \\ \hline
WD J0426$+$4044 & DA & 2 & 6395$\pm$94 & 7.98$\pm$0.03 & 1 & 7.69$^{+3.78}_{-2.91}$ \\ \hline
WD J0447$-$5823 & DA & 2 & 7630$\pm$110 & 7.99$\pm$0.03 & 1 & 6.95$^{+4.64}_{-3.04}$ \\ \hline 
WD J0525$-$1323 & DA & 1 & 5211$\pm$71 & 7.32$\pm$0.05 & 1 & $>$1.77$^{+0.12}_{-0.26}$ \\ \hline 
WD J0609$-$3035 & DA & 1 & 4848$\pm$103 & 7.86$\pm$0.06 & 1 & $>$4.21$^{+0.85}_{-0.61}$ \\ \hline
WD J0642$-$4721 & DA & 1 & 4450$\pm$163 & 7.63$\pm$0.11 & 1 & $>$3.66$^{+0.92}_{-0.62}$ \\ \hline
WD J0652$+$5628 & DA & 1 & 6048$\pm$196 & 8.16$\pm$0.07 & 1 & 4.69$^{+1.46}_{-0.66}$ & \\ \hline
WD J0716$+$0038 & {\it [DA]} & \dots & 4946$\pm$450 & 7.80$\pm$0.43 & 8 & $>$3.79$^{+4.55}_{-1.70}$ \\
 & {\it [non-DA]} & \dots & 4925$\pm$380 & 7.82$\pm$0.38 & 8 & $>$4.21$^{+2.90}_{-1.86}$ \\ \hline
WD J0735$-$7944 & DA & 2 & 28063$\pm$836 & 8.86$\pm$0.03 & 2 & 0.21$^{+0.02}_{-0.02}$ & \\ \hline
WD J0806$+$2215 & DA & 4 & 5760$\pm$35 & 8.07$\pm$0.02 & 9 & 5.56$^{+3.08}_{-0.98}$ \\ \hline
WD J0807$-$5350 & DA & 1 & 4433$\pm$125 & 7.53$\pm$0.09 & 1 & $>$3.25$^{+0.57}_{-0.40}$ \\ \hline
WD J0813$-$1358 & non-DA & 2 & 6451$\pm$131 & 7.97$\pm$0.04 & 1 & 8.71$^{+3.36}_{-3.15}$ \\ \hline
WD J0824$-$2546 & DA & 1 & 5081$\pm$136 & 7.92$\pm$0.08 & 1 & 10.60$^{+2.85}_{-2.50}$ \\ \hline 
WD J0950$+$1309 & DA & 1 & 5968$\pm$85 & 8.20$\pm$0.03 & 1 & 4.64$^{+0.40}_{-0.33}$ \\ 
 & non-DA & 2 & 5770$\pm$128 & 8.16$\pm$0.06 & 2 & 6.10$^{+2.32}_{-0.77}$ \\ \hline
WD J1042$+$3422 & DA: & 1 & 5510$\pm$155 & 8.58$\pm$0.07 & 1 & 7.82$^{+0.69}_{-0.68}$ \\ \hline
WD J1046$+$6054 & DC: & 1 & 5162$\pm$39 & 8.00$\pm$0.02 & 1 & 10.81$^{+2.34}_{-2.33}$ \\ \hline
WD J1125$-$1631 & DA & 1 & 6514$\pm$25 & 7.06$\pm$0.01 & 1 & $>$0.63$^{+0.02}_{-0.01}$ \\ \hline
WD J1157$+$0631 & DA & 5 & 7871$\pm$26 & 7.49$\pm$0.08 & 7 & $>$0.69$^{+0.05}_{-0.06}$ \\ \hline
WD J1234$+$4427 & DA & 1 & 5384$\pm$126 & 8.11$\pm$0.07 & 1 & 7.17$^{+2.81}_{-1.18}$ & \\ \hline
WD J1250$+$4152 & {\it [DA]} & \dots & 4060$\pm$680 & 7.84$\pm$0.66 & 10 & $>$7.53$^{+3.69}_{-4.17}$ \\ 
 & {\it [non-DA]} & \dots & 4098$\pm$375 & 7.78$\pm$0.56 & 8 & $>$6.59$^{+2.01}_{-3.05}$ \\ \hline
WD J1344$-$5738 & DA & 1 & 5733$\pm$88 & 7.79$\pm$0.04 & 1 & $>$2.22$^{+0.27}_{-0.28}$ \\ \hline 
WD J1418$+$0929 & DA & 6 & 15572$\pm$149 & 7.93$\pm$0.01 & 1 & 7.08$^{+4.18}_{-3.37}$ \\ \hline 
WD J1421$+$1436 & DA & 7 & 5407$\pm$99 & 7.27$\pm$0.25 & 7 & $>$1.58$^{+0.44}_{-0.30}$ \\ \hline
WD J1505$+$3304 & DA: & 1 & 4299$\pm$178 & 7.62$\pm$0.12 & 1 & $>$4.01$^{+0.96}_{-0.72}$ \\ \hline
WD J1628$-$0418 & \dots & \dots & \dots & \dots & \dots & \dots \\ \hline
WD J1642$+$5151 & {\it [DA]} & \dots & 5125$\pm$419 & 8.21$\pm$0.33 & 8 & 9.08$^{+2.85}_{-2.27}$ \\ 
 & {\it [non-DA]} & \dots & 5020$\pm$377 & 8.14$\pm$0.32 & 8 & 8.79$^{+3.44}_{-1.56}$ \\ \hline
WD J1657$-$1256 & DA & 1 & \dots & \dots & \dots & \dots \\ \hline
WD J1715$-$1910 & DA: & 1 & 4755$\pm$110 & 7.93$\pm$0.06 & 1 & 11.83$^{+2.10}_{-2.08}$ \\ \hline
WD J2005$+$6932 & DA: & 1 & 4937$\pm$99 & 8.02$\pm$0.05 & 1 & 10.54$^{+2.72}_{-1.86}$ \\ \hline
WD J2025$+$3515 & DA & 2 & 8293$\pm$105 & 8.02$\pm$0.02 & 1 & 5.35$^{+4.70}_{-2.32}$ \\ \hline
WD J2051$-$3558 & DA & 2 & 9053$\pm$114 & 8.23$\pm$0.02 & 1 & 1.68$^{+0.22}_{-0.11}$ \\ \hline
WD J2053$+$2603 & \dots & \dots & \dots & \dots & \dots & \dots \\ \hline
WD J2119$+$4206 & DA & 2 & 11691$\pm$111 & 7.81$\pm$0.01 & 2 & $>$0.32$^{+0.01}_{-0.01}$ \\ \hline
WD J2339$+$2552 & DA & 1 & 7489$\pm$47 & 6.88$\pm$0.02 & 1 & \dots \\ \hline
WD J2353$-$4325 & {\it [DA]} & \dots & 4427$\pm$438 & 8.16$\pm$0.36 & 8 & 11.13$^{+2.03}_{-1.92}$ \\ 
 & {\it [non-DA]} & \dots & 4422$\pm$343 & 8.10$\pm$0.35 & 8 & 9.92$^{+2.90}_{-1.27}$ \\ \hline
WD J2355$+$5131 & DA & 1 & 5706$\pm$151 & 8.42$\pm$0.06 & 1 & 6.53$^{+0.71}_{-0.59}$ \\ \hline 
WD J2357$+$3043 & DA: & 1 & 5997$\pm$67 & 7.35$\pm$0.05 & 1 & $>$1.24$^{+0.09}_{-0.11}$ \\ \hline
\enddata
\tablecomments{For objects with surface gravities outside of the lower limit of 7.9 set in \cite{kiman2022}, the cooling age is treated as a lower age limit.}
\tablerefs{(1) \cite{vincent2024}; (2) \cite{jimenez2023}; (3) \cite{kleinman2004}; (4) \cite{limoges2015}; (5) \cite{girven2011}; (6) \cite{brown2007}; (7) \cite{kepler2019}; (8) \cite{gentile2021}; (9) \cite{blouin2019}; (10) \cite{gentile2019}; (11) This work.  }
\end{deluxetable*}

\section{Conclusions}
\label{sec:conclusion}
In this study, we identified and compiled a new set of ultracool dwarf companions to white dwarfs within 100 pc using available astrometry and optical and infrared data. Through our search, we identified fifty-one previously unrecognized systems with candidate ultracool companions. Of these, thirty-one are resolved in at least one catalogue, with all but six reliably confirmed as co-moving based on consistent proper motion and parallax measurements (if available). The six unreliable cases will require additional astrometric data to verify their status as companions to the white dwarf primary. 

Follow-up near-infrared spectroscopy of four of our resolved co-moving candidates confirmed their ultracool nature with three confirmed to be M8, and one a likely stellar source with an L3 spectral type.  However, one of these systems (WD J2119$+$4206) was deemed unlikely to be physically associated in our astrometric analysis.  Three of our unresolved sources were confirmed as having cool stellar companions via spectral decomposition of available optical spectra from Gaia. The remaining candidates will benefit from detailed follow-up spectroscopy to further constrain the natures of these systems.

\section{Acknowledgements}
The authors acknowledge support from the Science and Engineering Apprenticeship Program (SEAP) of the Office of Naval Research. 

This publication makes use of data products from the {\it Wide-field Infrared Survey Explorer}, which is a joint project of the University of California, Los Angeles, and the Jet Propulsion Laboratory/California Institute of Technology, and NEOWISE which is a project of the Jet Propulsion Laboratory/California Institute of Technology. {\it WISE} and NEOWISE are funded by the National Aeronautics and Space Administration. Part of this research was carried out at the Jet Propulsion Laboratory, California Institute of Technology, under a contract with the National Aeronautics and Space Administration. 

This publication makes use of data products from the UKIRT Hemisphere Survey, which is a joint project of the United States Naval Observatory, the University of Hawaii Institute for Astronomy, the Cambridge University Cambridge Astronomy Survey Unit, and the University of Edinburgh Wide-Field Astronomy Unit (WFAU). This project was primarily funded by the United States Navy. The WFAU gratefully acknowledges support for this work from the Science and Technology Facilities Council through ST/T002956/1 and previous grants. 

The Pan-STARRS1 Surveys (PS1) and the PS1 public science archive have been made possible through contributions by the Institute for Astronomy, the University of Hawaii, the Pan-STARRS Project Office, the Max-Planck Society and its participating institutes, the Max Planck Institute for Astronomy, Heidelberg and the Max Planck Institute for Extraterrestrial Physics, Garching, The Johns Hopkins University, Durham University, the University of Edinburgh, the Queen's University Belfast, the Harvard-Smithsonian Center for Astrophysics, the Las Cumbres Observatory Global Telescope Network Incorporated, the National Central University of Taiwan, the Space Telescope Science Institute, the National Aeronautics and Space Administration under Grant No. NNX08AR22G issued through the Planetary Science Division of the NASA Science Mission Directorate, the National Science Foundation Grant No. AST-1238877, the University of Maryland, Eotvos Lorand University (ELTE), the Los Alamos National Laboratory, and the Gordon and Betty Moore Foundation.

This publication has made use of the Python package GaiaXPy, developed and maintained by members of the Gaia Data Processing and Analysis Consortium (DPAC), and in particular, Coordination Unit 5 (CU5), and the Data Processing Centre located at the Institute of Astronomy, Cambridge, UK (DPCI).

This research uses services or data provided by the Astro Data Lab, which is part of the Community Science and Data Center (CSDC) Program of NSF NOIRLab. NOIRLab is operated by the Association of Universities for Research in Astronomy (AURA), Inc. under a cooperative agreement with the U.S. National Science Foundation.

This work has benefitted from The UltracoolSheet at http://bit.ly/UltracoolSheet, maintained by Will Best, Trent Dupuy, Michael Liu, Aniket Sanghi, Rob Siverd, and Zhoujian Zhang, and developed from compilations by \cite{dupuy2012}, \cite{dupuy2013}, \cite{liu2016}, \cite{best2018}, \cite{best2021}, \cite{sanghi2023}, and \cite{schneider2023}.

\software{\texttt{CoMover} \citep{gagne2021}; GaiaXPy \citep{ruz2024}; SpexTool \citep{cushing2004}; WiseView \citep{caselden2018}} 
\facilities{Astro Data Lab, IRTF} 

\appendix

\section{White Dwarf-White Dwarf Systems}
\label{sec:appA}
Our search for cold companions using Gaia (Method III in Section \ref{sec:cans}) identified many likely white-dwarf-white dwarf systems.  Listed in Table \ref{tab:wdwd} are the 115 potential white dwarf-white dwarf systems found through our search where at least one component in the system has an absolute $G$-band magnitude $>$14.5 mag.

\startlongtable
\begin{deluxetable*}{ccccccccccccccc}
\tabletypesize{\footnotesize}
\tablecaption{Candidate White Dwarf-White Dwarf Systems}
\label{tab:wdwd}
\tablehead{\colhead{WD Name}&\colhead{R.A.}&\colhead{Dec.}&\colhead{Sep.}&\colhead{$\varpi$}&\colhead{$\mu_{\alpha}$}&\colhead{$\mu_{\delta}$}\\
&\colhead{($\degr$)}&\colhead{($\degr$)}&\colhead{($\arcsec$)}&\colhead{(mas)}&\colhead{(mas yr$^{-1}$)}&\colhead{(mas yr$^{-1}$)}}
 \startdata
WD J001440.69$-$075856.97	 & 	3.6703727	 & 	-7.9824549	 & 		 & 	10.7067	$\pm$	0.2773	 & 	184.026	$\pm$	0.304	 & 	8.642	$\pm$	0.221	 \\	
WD J001440.94$-$075857.72	 & 	3.6714185	 & 	-7.9826578	 & 	3.799	 & 	10.6252	$\pm$	0.3461	 & 	184.174	$\pm$	0.378	 & 	10.656	$\pm$	0.282	 \\	\hline
WD J002229.32$-$723209.63	 & 	5.6244136	 & 	-72.5357083	 & 		 & 	11.2515	$\pm$	0.0882	 & 	152.512	$\pm$	0.121	 & 	67.550	$\pm$	0.098	 \\	
WD J002232.67$-$723233.64	 & 	5.6383606	 & 	-72.5423760	 & 	28.340	 & 	11.1783	$\pm$	0.2545	 & 	152.169	$\pm$	0.388	 & 	67.094	$\pm$	0.298	 \\	\hline
WD J004920.31$-$833350.65	 & 	12.3369484	 & 	-83.5649155	 & 		 & 	17.9326	$\pm$	0.1528	 & 	59.305	$\pm$	0.190	 & 	-190.426	$\pm$	0.181	 \\	
WD J004922.35$-$833352.55	 & 	12.3454283	 & 	-83.5654492	 & 	3.924	 & 	18.1349	$\pm$	0.1394	 & 	57.911	$\pm$	0.168	 & 	-191.591	$\pm$	0.169	 \\	\hline
WD J005133.15$-$671516.04	 & 	12.8890978	 & 	-67.2551668	 & 		 & 	11.1938	$\pm$	0.2352	 & 	85.748	$\pm$	0.288	 & 	-160.402	$\pm$	0.283	 \\	
WD J005134.57$-$671517.80	 & 	12.8950200	 & 	-67.2556566	 & 	8.429	 & 	11.3066	$\pm$	0.3234	 & 	85.533	$\pm$	0.376	 & 	-159.316	$\pm$	0.388	 \\	\hline
WD J005550.32$+$854500.21	 & 	13.9825902	 & 	85.7509450	 & 		 & 	15.3529	$\pm$	0.2100	 & 	381.124	$\pm$	0.247	 & 	199.250	$\pm$	0.244	 \\	
WD J005551.69$+$854509.76	 & 	13.9883628	 & 	85.7535899	 & 	9.645	 & 	15.2162	$\pm$	0.2253	 & 	382.152	$\pm$	0.262	 & 	197.213	$\pm$	0.272	 \\	\hline
WD J010456.47$+$211958.87	 & 	16.2342953	 & 	21.3311062	 & 		 & 	30.6583	$\pm$	0.1101	 & 	-210.726	$\pm$	0.141	 & 	-430.405	$\pm$	0.095	 \\	
WD J010457.96$+$212017.54	 & 	16.2404882	 & 	21.3362579	 & 	27.843	 & 	30.8044	$\pm$	0.1138	 & 	-208.398	$\pm$	0.147	 & 	-438.580	$\pm$	0.101	 \\	\hline
WD J010903.42$-$104214.15	 & 	17.2650176	 & 	-10.7040573	 & 		 & 	16.4634	$\pm$	0.0666	 & 	173.007	$\pm$	0.095	 & 	-28.481	$\pm$	0.107	 \\	
WD J010904.22$-$104215.39	 & 	17.2683761	 & 	-10.7043972	 & 	11.943	 & 	16.7654	$\pm$	0.2253	 & 	174.414	$\pm$	0.366	 & 	-27.459	$\pm$	0.355	 \\	\hline
WD J011728.64$-$043939.68	 & 	19.3681657	 & 	-4.6622002	 & 		 & 	24.2006	$\pm$	0.1529	 & 	-264.474	$\pm$	0.209	 & 	-264.895	$\pm$	0.141	 \\	
WD J011728.83$-$043938.41	 & 	19.3689792	 & 	-4.6618450	 & 	3.187	 & 	23.5673	$\pm$	0.1535	 & 	-259.303	$\pm$	0.201	 & 	-264.617	$\pm$	0.142	 \\	\hline
WD J011911.55$+$312636.35	 & 	19.7971235	 & 	31.4429825	 & 		 & 	12.0849	$\pm$	0.3719	 & 	-190.959	$\pm$	0.368	 & 	-101.193	$\pm$	0.282	 \\	
WD J011911.80$+$312634.00	 & 	19.7981300	 & 	31.4423283	 & 	3.886	 & 	12.6415	$\pm$	0.3856	 & 	-195.632	$\pm$	0.381	 & 	-100.758	$\pm$	0.295	 \\	\hline
WD J012327.62$+$390532.81	 & 	20.8652823	 & 	39.0920769	 & 		 & 	11.4706	$\pm$	0.4602	 & 	36.599	$\pm$	0.359	 & 	-85.683	$\pm$	0.342	 \\	
WD J012328.11$+$390530.76	 & 	20.8673707	 & 	39.0914859	 & 	6.211	 & 	11.2825	$\pm$	0.5984	 & 	39.470	$\pm$	0.491	 & 	-88.154	$\pm$	0.450	 \\	\hline
WD J012621.66$-$611024.68	 & 	21.5916150	 & 	-61.1735373	 & 		 & 	12.4288	$\pm$	0.1153	 & 	149.823	$\pm$	0.111	 & 	-3.712	$\pm$	0.109	 \\	
WD J012622.26$-$611026.13	 & 	21.5941288	 & 	-61.1739247	 & 	4.581	 & 	12.3629	$\pm$	0.3444	 & 	148.097	$\pm$	0.320	 & 	0.649	$\pm$	0.348	 \\	\hline
WD J013445.35$+$642959.71	 & 	23.6903081	 & 	64.5004880	 & 		 & 	13.0164	$\pm$	0.2496	 & 	130.681	$\pm$	0.192	 & 	128.128	$\pm$	0.247	 \\	
WD J013445.69$+$642953.22	 & 	23.6917507	 & 	64.4986718	 & 	6.910	 & 	12.4739	$\pm$	0.2316	 & 	132.396	$\pm$	0.182	 & 	124.772	$\pm$	0.231	 \\	\hline
WD J015117.24$+$124106.19	 & 	27.8223269	 & 	12.6847248	 & 		 & 	10.7743	$\pm$	0.4056	 & 	103.307	$\pm$	0.483	 & 	-73.453	$\pm$	0.444	 \\	
WD J015117.40$+$124106.58	 & 	27.8229503	 & 	12.6848550	 & 	2.239	 & 	11.3616	$\pm$	0.4448	 & 	98.577	$\pm$	0.573	 & 	-71.234	$\pm$	0.480	 \\	\hline
WD J021255.73$-$475803.38	 & 	33.2322541	 & 	-47.9675755	 & 		 & 	12.5863	$\pm$	0.2404	 & 	8.070	$\pm$	0.238	 & 	6.429	$\pm$	0.249	 \\	
WD J021255.95$-$475803.61	 & 	33.2331681	 & 	-47.9676311	 & 	2.212	 & 	13.2548	$\pm$	0.2080	 & 	9.813	$\pm$	0.208	 & 	8.480	$\pm$	0.214	 \\	\hline
WD J022443.13$-$024255.30	 & 	36.1812965	 & 	-2.7149470	 & 		 & 	13.5405	$\pm$	0.2844	 & 	359.925	$\pm$	0.341	 & 	92.780	$\pm$	0.249	 \\	
WD J022443.44$-$024257.96	 & 	36.1825989	 & 	-2.7156982	 & 	5.408	 & 	13.4107	$\pm$	0.2742	 & 	358.466	$\pm$	0.316	 & 	89.655	$\pm$	0.229	 \\	\hline
WD J022551.96$+$422803.52	 & 	36.4688877	 & 	42.4667255	 & 		 & 	19.7947	$\pm$	0.0788	 & 	397.472	$\pm$	0.091	 & 	-206.848	$\pm$	0.086	 \\	
WD J022552.59$+$422803.52	 & 	36.4715238	 & 	42.4667227	 & 	7.001	 & 	20.1056	$\pm$	0.2148	 & 	399.941	$\pm$	0.258	 & 	-207.737	$\pm$	0.239	 \\	\hline
WD J023258.11$-$664822.91	 & 	38.2423462	 & 	-66.8066161	 & 		 & 	17.0207	$\pm$	0.1038	 & 	21.131	$\pm$	0.119	 & 	-56.449	$\pm$	0.133	 \\	
WD J023258.54$-$664817.55	 & 	38.2441728	 & 	-66.8051491	 & 	5.882	 & 	17.2730	$\pm$	0.1444	 & 	23.430	$\pm$	0.163	 & 	-61.066	$\pm$	0.185	 \\	\hline
WD J024611.55$-$265128.81	 & 	41.5490878	 & 	-26.8573504	 & 		 & 	11.1575	$\pm$	0.2104	 & 	191.524	$\pm$	0.154	 & 	146.765	$\pm$	0.196	 \\	
WD J024612.62$-$265136.90	 & 	41.5535529	 & 	-26.8596060	 & 	16.480	 & 	12.0314	$\pm$	0.5176	 & 	191.239	$\pm$	0.384	 & 	146.596	$\pm$	0.550	 \\	\hline
WD J025910.80$-$045531.15	 & 	44.7945286	 & 	-4.9258500	 & 		 & 	10.4737	$\pm$	0.4913	 & 	-104.082	$\pm$	0.504	 & 	-118.864	$\pm$	0.530	 \\	
WD J025910.81$-$045532.99	 & 	44.7945872	 & 	-4.9263602	 & 	1.848	 & 	12.5394	$\pm$	0.4345	 & 	-99.357	$\pm$	0.450	 & 	-119.470	$\pm$	0.488	 \\	\hline
WD J031945.74$+$640904.32	 & 	49.9426141	 & 	64.1510268	 & 		 & 	20.5635	$\pm$	0.1207	 & 	199.488	$\pm$	0.071	 & 	-38.548	$\pm$	0.106	 \\	
WD J031946.40$+$640854.18	 & 	49.9453785	 & 	64.1482168	 & 	11.007	 & 	20.5931	$\pm$	0.1454	 & 	200.453	$\pm$	0.088	 & 	-37.170	$\pm$	0.132	 \\	\hline
WD J032414.19$+$611350.38	 & 	51.0586146	 & 	61.2304012	 & 		 & 	12.6170	$\pm$	0.2636	 & 	-54.220	$\pm$	0.285	 & 	-58.499	$\pm$	0.315	 \\	
WD J032414.66$+$611347.71	 & 	51.0606110	 & 	61.2296680	 & 	4.351	 & 	11.8883	$\pm$	0.2757	 & 	-50.591	$\pm$	0.295	 & 	-56.967	$\pm$	0.331	 \\	\hline
WD J033304.60$-$563833.16	 & 	53.2694388	 & 	-56.6424918	 & 		 & 	10.6104	$\pm$	0.0413	 & 	33.987	$\pm$	0.048	 & 	12.067	$\pm$	0.049	 \\	
WD J033305.06$-$563833.52	 & 	53.2713546	 & 	-56.6425999	 & 	3.812	 & 	10.4286	$\pm$	0.2843	 & 	32.774	$\pm$	0.368	 & 	10.431	$\pm$	0.342	 \\	\hline
WD J033649.70$+$641707.31	 & 	54.2080310	 & 	64.2846282	 & 		 & 	13.7655	$\pm$	0.2684	 & 	92.680	$\pm$	0.147	 & 	-164.991	$\pm$	0.240	 \\	
WD J033650.93$+$641709.86	 & 	54.2131407	 & 	64.2853412	 & 	8.384	 & 	13.8713	$\pm$	0.1892	 & 	91.735	$\pm$	0.102	 & 	-164.971	$\pm$	0.174	 \\	\hline
WD J034501.53$-$034849.73	 & 	56.2563468	 & 	-3.8149903	 & 		 & 	32.2247	$\pm$	0.1154	 & 	-8.806	$\pm$	0.102	 & 	-264.757	$\pm$	0.102	 \\	
WD J034501.70$-$034844.85	 & 	56.2569717	 & 	-3.8136270	 & 	5.397	 & 	32.1842	$\pm$	0.0820	 & 	-28.531	$\pm$	0.074	 & 	-263.147	$\pm$	0.072	 \\	\hline
WD J035012.13$-$383057.28	 & 	57.5505914	 & 	-38.5159057	 & 		 & 	11.3036	$\pm$	0.1698	 & 	11.674	$\pm$	0.160	 & 	1.050	$\pm$	0.210	 \\	
WD J035012.19$-$383051.98	 & 	57.5508536	 & 	-38.5144322	 & 	5.356	 & 	10.9623	$\pm$	0.2555	 & 	10.929	$\pm$	0.243	 & 	1.314	$\pm$	0.329	 \\	\hline
WD J035656.96$+$521345.45	 & 	59.2376134	 & 	52.2285303	 & 		 & 	10.9369	$\pm$	0.2482	 & 	39.435	$\pm$	0.260	 & 	-171.191	$\pm$	0.283	 \\	
WD J035657.13$+$521345.97	 & 	59.2383243	 & 	52.2286519	 & 	1.628	 & 	11.6414	$\pm$	0.2580	 & 	37.653	$\pm$	0.277	 & 	-175.117	$\pm$	0.283	 \\	\hline
WD J040229.20$+$481256.57	 & 	60.6225797	 & 	48.2147447	 & 		 & 	21.3730	$\pm$	0.1226	 & 	138.754	$\pm$	0.130	 & 	-218.119	$\pm$	0.106	 \\	
WD J040229.40$+$481256.62	 & 	60.6234670	 & 	48.2147260	 & 	2.129	 & 	21.6579	$\pm$	0.1272	 & 	142.673	$\pm$	0.142	 & 	-225.472	$\pm$	0.110	 \\	\hline
WD J040952.26$+$195244.53	 & 	62.4677361	 & 	19.8786747	 & 		 & 	12.0577	$\pm$	0.5616	 & 	-1.153	$\pm$	0.513	 & 	-81.357	$\pm$	0.464	 \\	
WD J040952.71$+$195249.15	 & 	62.4696342	 & 	19.8799715	 & 	7.943	 & 	10.5242	$\pm$	0.3761	 & 	-0.800	$\pm$	0.388	 & 	-79.754	$\pm$	0.347	 \\	\hline
WD J041612.00$-$822323.42	 & 	64.0519086	 & 	-82.3893146	 & 		 & 	17.3770	$\pm$	0.1672	 & 	57.846	$\pm$	0.234	 & 	118.378	$\pm$	0.207	 \\	
WD J041612.38$-$822326.26	 & 	64.0537485	 & 	-82.3900748	 & 	2.874	 & 	17.4722	$\pm$	0.1765	 & 	64.543	$\pm$	0.235	 & 	123.504	$\pm$	0.218	 \\	\hline
WD J042029.21$+$261755.39	 & 	65.1213554	 & 	26.2985880	 & 		 & 	11.2535	$\pm$	0.1570	 & 	-69.754	$\pm$	0.201	 & 	-29.354	$\pm$	0.151	 \\	
WD J042031.95$+$261708.87	 & 	65.1327895	 & 	26.2856696	 & 	59.370	 & 	11.6084	$\pm$	0.3722	 & 	-68.510	$\pm$	0.477	 & 	-28.589	$\pm$	0.355	 \\	\hline
WD J042535.48$+$021009.86	 & 	66.3983522	 & 	2.1695532	 & 		 & 	17.2978	$\pm$	0.1930	 & 	114.372	$\pm$	0.200	 & 	33.879	$\pm$	0.158	 \\	
WD J042535.86$+$021009.42	 & 	66.3999003	 & 	2.1694361	 & 	5.585	 & 	16.7394	$\pm$	0.2105	 & 	112.336	$\pm$	0.215	 & 	34.463	$\pm$	0.170	 \\	\hline
WD J043726.68$+$291545.16	 & 	69.3615629	 & 	29.2621674	 & 		 & 	16.1993	$\pm$	0.2221	 & 	79.593	$\pm$	0.277	 & 	-85.011	$\pm$	0.190	 \\	
WD J043728.64$+$291522.49	 & 	69.3697597	 & 	29.2558761	 & 	34.289	 & 	16.5762	$\pm$	0.2266	 & 	79.647	$\pm$	0.285	 & 	-83.509	$\pm$	0.196	 \\	\hline
WD J044457.95$+$093401.24	 & 	71.2417748	 & 	9.5669396	 & 		 & 	11.4687	$\pm$	0.5674	 & 	69.953	$\pm$	0.671	 & 	-15.633	$\pm$	0.445	 \\	
WD J044458.06$+$093403.60	 & 	71.2422082	 & 	9.5675986	 & 	2.828	 & 	11.3038	$\pm$	0.5702	 & 	67.375	$\pm$	0.661	 & 	-16.704	$\pm$	0.416	 \\	\hline
WD J050517.30$+$192851.34	 & 	76.3230799	 & 	19.4805376	 & 		 & 	19.3568	$\pm$	0.1995	 & 	215.007	$\pm$	0.226	 & 	-87.868	$\pm$	0.160	 \\	
WD J050517.64$+$192844.93	 & 	76.3245093	 & 	19.4787709	 & 	7.999	 & 	19.6058	$\pm$	0.1876	 & 	215.292	$\pm$	0.214	 & 	-84.775	$\pm$	0.152	 \\	\hline
WD J050640.18$-$551052.40	 & 	76.6682322	 & 	-55.1805613	 & 		 & 	10.2198	$\pm$	0.0635	 & 	102.601	$\pm$	0.076	 & 	148.630	$\pm$	0.085	 \\	
WD J050641.25$-$551129.39	 & 	76.6726869	 & 	-55.1908311	 & 	38.088	 & 	10.2629	$\pm$	0.2738	 & 	102.844	$\pm$	0.344	 & 	147.837	$\pm$	0.408	 \\	\hline
WD J051250.01$+$473534.63	 & 	78.2080629	 & 	47.5909691	 & 		 & 	17.3859	$\pm$	0.2164	 & 	-46.528	$\pm$	0.267	 & 	-446.287	$\pm$	0.232	 \\	
WD J051251.03$+$473545.19	 & 	78.2123307	 & 	47.5939120	 & 	14.819	 & 	17.6029	$\pm$	0.3755	 & 	-47.415	$\pm$	0.458	 & 	-444.349	$\pm$	0.402	 \\	\hline
WD J054705.48$+$753139.63	 & 	86.7755098	 & 	75.5275867	 & 		 & 	13.5239	$\pm$	0.1874	 & 	150.647	$\pm$	0.175	 & 	-20.102	$\pm$	0.210	 \\	
WD J054706.58$+$753103.10	 & 	86.7800858	 & 	75.5174347	 & 	36.778	 & 	13.2976	$\pm$	0.1880	 & 	149.399	$\pm$	0.165	 & 	-19.724	$\pm$	0.205	 \\	\hline
WD J061209.21$-$734313.20	 & 	93.0389360	 & 	-73.7210593	 & 		 & 	11.3281	$\pm$	0.2043	 & 	34.640	$\pm$	0.231	 & 	-162.724	$\pm$	0.233	 \\	
WD J061209.44$-$734311.17	 & 	93.0398145	 & 	-73.7204831	 & 	2.256	 & 	11.4147	$\pm$	0.2964	 & 	31.449	$\pm$	0.347	 & 	-161.964	$\pm$	0.333	 \\	\hline
WD J061806.18$-$110537.57	 & 	94.5249766	 & 	-11.0941059	 & 		 & 	16.8912	$\pm$	0.2176	 & 	-169.017	$\pm$	0.210	 & 	-75.674	$\pm$	0.248	 \\	
WD J061806.22$-$110541.87	 & 	94.5251738	 & 	-11.0952995	 & 	4.353	 & 	16.8600	$\pm$	0.1811	 & 	-163.065	$\pm$	0.171	 & 	-75.551	$\pm$	0.209	 \\	\hline
WD J063347.72$+$341532.56	 & 	98.4486852	 & 	34.2592127	 & 		 & 	10.7504	$\pm$	0.5436	 & 	-30.148	$\pm$	0.545	 & 	37.643	$\pm$	0.458	 \\	
WD J063348.21$+$341536.93	 & 	98.4507325	 & 	34.2604230	 & 	7.489	 & 	11.1398	$\pm$	0.3988	 & 	-29.361	$\pm$	0.392	 & 	36.116	$\pm$	0.330	 \\	\hline
WD J070326.09$+$622251.24	 & 	105.8576427	 & 	62.3799757	 & 		 & 	12.2384	$\pm$	0.0342	 & 	-110.722	$\pm$	0.033	 & 	-208.158	$\pm$	0.030	 \\	
WD J070326.56$+$622253.41	 & 	105.8596145	 & 	62.3805775	 & 	3.940	 & 	11.2593	$\pm$	0.5696	 & 	-109.421	$\pm$	0.546	 & 	-208.207	$\pm$	0.494	 \\	\hline
WD J071038.98$-$644643.37	 & 	107.6621750	 & 	-64.7775440	 & 		 & 	11.9505	$\pm$	0.1501	 & 	-23.050	$\pm$	0.224	 & 	263.060	$\pm$	0.164	 \\	
WD J071039.37$-$644646.89	 & 	107.6637879	 & 	-64.7785290	 & 	4.324	 & 	12.0441	$\pm$	0.2487	 & 	-22.418	$\pm$	0.370	 & 	261.238	$\pm$	0.279	 \\	\hline
WD J074546.92$-$825400.39	 & 	116.4462945	 & 	-82.8996420	 & 		 & 	10.3798	$\pm$	0.3848	 & 	20.152	$\pm$	0.439	 & 	104.870	$\pm$	0.379	 \\	
WD J074548.49$-$825402.30	 & 	116.4528923	 & 	-82.9001779	 & 	3.513	 & 	10.1995	$\pm$	0.3511	 & 	22.847	$\pm$	0.406	 & 	104.988	$\pm$	0.358	 \\	\hline
WD J075014.58$+$071148.92	 & 	117.5617156	 & 	7.1889986	 & 		 & 	54.9675	$\pm$	0.0611	 & 	211.520	$\pm$	0.071	 & 	-1782.618	$\pm$	0.047	 \\	
WD J075015.34$+$071137.10	 & 	117.5648647	 & 	7.1856806	 & 	16.407	 & 	55.1048	$\pm$	0.0557	 & 	209.910	$\pm$	0.068	 & 	-1790.496	$\pm$	0.044	 \\	\hline
WD J075442.93$-$032224.13	 & 	118.6781482	 & 	-3.3731744	 & 		 & 	15.7229	$\pm$	0.3192	 & 	-160.882	$\pm$	0.309	 & 	44.037	$\pm$	0.223	 \\	
WD J075443.04$-$032228.99	 & 	118.6786196	 & 	-3.3745372	 & 	5.191	 & 	16.0783	$\pm$	0.3182	 & 	-157.463	$\pm$	0.297	 & 	40.750	$\pm$	0.214	 \\	\hline
WD J080729.69$-$391206.96	 & 	121.8739434	 & 	-39.2018902	 & 		 & 	13.5017	$\pm$	0.2196	 & 	37.888	$\pm$	0.228	 & 	10.302	$\pm$	0.278	 \\	
WD J080730.53$-$391155.69	 & 	121.8774103	 & 	-39.1987706	 & 	14.821	 & 	13.3886	$\pm$	0.2053	 & 	37.828	$\pm$	0.207	 & 	7.570	$\pm$	0.264	 \\	\hline
WD J082944.78$-$525705.69	 & 	127.4368508	 & 	-52.9519422	 & 		 & 	13.5560	$\pm$	0.5197	 & 	38.460	$\pm$	0.627	 & 	-81.206	$\pm$	0.662	 \\	
WD J082944.92$-$525707.51	 & 	127.4374067	 & 	-52.9524223	 & 	2.107	 & 	12.7794	$\pm$	0.1816	 & 	32.774	$\pm$	0.218	 & 	-76.525	$\pm$	0.221	 \\	\hline
WD J083105.19$-$765253.79	 & 	127.7734957	 & 	-76.8820559	 & 		 & 	13.1626	$\pm$	0.2610	 & 	96.458	$\pm$	0.348	 & 	-101.377	$\pm$	0.283	 \\	
WD J083105.82$-$765251.63	 & 	127.7761478	 & 	-76.8814513	 & 	3.071	 & 	12.3257	$\pm$	0.2441	 & 	95.969	$\pm$	0.363	 & 	-99.739	$\pm$	0.273	 \\	\hline
WD J092551.68$+$354000.58	 & 	141.4653750	 & 	35.6663662	 & 		 & 	14.4284	$\pm$	0.1149	 & 	8.699	$\pm$	0.140	 & 	-103.763	$\pm$	0.153	 \\	
WD J092551.76$+$353957.73	 & 	141.4657031	 & 	35.6655539	 & 	3.078	 & 	14.3515	$\pm$	0.2314	 & 	10.340	$\pm$	0.294	 & 	-108.278	$\pm$	0.312	 \\	\hline
WD J094722.99$+$445948.59	 & 	146.8462272	 & 	44.9970255	 & 		 & 	18.5394	$\pm$	0.2179	 & 	72.654	$\pm$	0.203	 & 	43.308	$\pm$	0.162	 \\	
WD J094724.45$+$450001.85	 & 	146.8523191	 & 	45.0007018	 & 	20.388	 & 	18.1471	$\pm$	0.1914	 & 	71.556	$\pm$	0.176	 & 	42.095	$\pm$	0.143	 \\	\hline
WD J100553.29$-$193056.55	 & 	151.4709362	 & 	-19.5152295	 & 		 & 	11.0226	$\pm$	0.5048	 & 	-238.087	$\pm$	0.490	 & 	107.920	$\pm$	0.570	 \\	
WD J100553.56$-$193055.83	 & 	151.4720682	 & 	-19.5150358	 & 	3.904	 & 	11.6713	$\pm$	0.5403	 & 	-235.593	$\pm$	0.541	 & 	105.383	$\pm$	0.672	 \\	\hline
WD J100623.08$+$071212.70	 & 	151.5963157	 & 	7.2032200	 & 		 & 	18.3308	$\pm$	0.0583	 & 	37.925	$\pm$	0.061	 & 	-69.130	$\pm$	0.059	 \\	
WD J100623.17$+$071154.30	 & 	151.5967276	 & 	7.1981046	 & 	18.474	 & 	18.1518	$\pm$	0.2255	 & 	37.308	$\pm$	0.259	 & 	-70.214	$\pm$	0.264	 \\	\hline
WD J101148.82$+$464929.83	 & 	152.9532581	 & 	46.8246483	 & 		 & 	10.8350	$\pm$	0.3076	 & 	-25.295	$\pm$	0.269	 & 	-68.902	$\pm$	0.272	 \\	
WD J101148.83$+$464924.17	 & 	152.9532848	 & 	46.8230704	 & 	5.681	 & 	10.8824	$\pm$	0.5624	 & 	-23.352	$\pm$	0.496	 & 	-70.025	$\pm$	0.540	 \\	\hline
WD J101359.85$+$030553.90	 & 	153.4998209	 & 	3.0978670	 & 		 & 	20.1351	$\pm$	0.2443	 & 	105.073	$\pm$	0.184	 & 	-99.129	$\pm$	0.213	 \\	
WD J101401.60$+$030550.42	 & 	153.5071480	 & 	3.0968880	 & 	26.574	 & 	20.6894	$\pm$	0.1936	 & 	110.107	$\pm$	0.190	 & 	-101.150	$\pm$	0.227	 \\	\hline
WD J101539.25$-$752747.50	 & 	153.9110633	 & 	-75.4634967	 & 		 & 	11.2690	$\pm$	0.2129	 & 	-139.455	$\pm$	0.311	 & 	-67.561	$\pm$	0.274	 \\	
WD J101539.82$-$752745.82	 & 	153.9134274	 & 	-75.4630451	 & 	2.684	 & 	11.5409	$\pm$	0.4828	 & 	-140.339	$\pm$	0.705	 & 	-71.243	$\pm$	0.656	 \\	\hline
WD J102442.01$+$491351.28	 & 	156.1737922	 & 	49.2304852	 & 		 & 	12.6014	$\pm$	0.2865	 & 	-181.958	$\pm$	0.210	 & 	-95.849	$\pm$	0.313	 \\	
WD J102442.57$+$491407.82	 & 	156.1761480	 & 	49.2350740	 & 	17.423	 & 	11.5901	$\pm$	0.4260	 & 	-182.426	$\pm$	0.344	 & 	-96.637	$\pm$	0.443	 \\	\hline
WD J103002.22$-$310135.20	 & 	157.5087674	 & 	-31.0267345	 & 		 & 	10.2653	$\pm$	0.4845	 & 	-89.626	$\pm$	0.339	 & 	-64.798	$\pm$	0.474	 \\	
WD J103002.31$-$310138.67	 & 	157.5091676	 & 	-31.0276912	 & 	3.659	 & 	10.2087	$\pm$	0.4990	 & 	-85.395	$\pm$	0.380	 & 	-65.254	$\pm$	0.493	 \\	\hline
WD J112952.04$-$312247.21	 & 	172.4652769	 & 	-31.3788066	 & 		 & 	13.8999	$\pm$	0.1652	 & 	-296.162	$\pm$	0.191	 & 	219.728	$\pm$	0.122	 \\	
WD J112953.77$-$312243.02	 & 	172.4724986	 & 	-31.3776397	 & 	22.590	 & 	14.2005	$\pm$	0.3007	 & 	-297.061	$\pm$	0.341	 & 	219.713	$\pm$	0.219	 \\	\hline
WD J122302.62$-$340800.00	 & 	185.7599176	 & 	-34.1339639	 & 		 & 	11.8074	$\pm$	0.6166	 & 	-188.470	$\pm$	0.752	 & 	-141.658	$\pm$	0.584	 \\	
WD J122302.68$-$340802.31	 & 	185.7601731	 & 	-34.1345908	 & 	2.382	 & 	13.4927	$\pm$	0.7096	 & 	-186.779	$\pm$	0.895	 & 	-138.925	$\pm$	0.796	 \\	\hline
WD J124029.35$+$255948.30	 & 	190.1211382	 & 	25.9965929	 & 		 & 	10.2865	$\pm$	0.5681	 & 	-235.037	$\pm$	0.627	 & 	-35.049	$\pm$	0.491	 \\	
WD J124029.56$+$255947.81	 & 	190.1220258	 & 	25.9964738	 & 	2.904	 & 	10.2112	$\pm$	0.4076	 & 	-233.955	$\pm$	0.531	 & 	-32.550	$\pm$	0.349	 \\	\hline
WD J124824.48$-$664037.75	 & 	192.0983149	 & 	-66.6772751	 & 		 & 	13.6669	$\pm$	0.3154	 & 	-327.645	$\pm$	0.259	 & 	-27.728	$\pm$	0.305	 \\	
WD J124825.45$-$664013.50	 & 	192.1023772	 & 	-66.6705433	 & 	24.917	 & 	14.5679	$\pm$	0.5394	 & 	-328.029	$\pm$	0.468	 & 	-28.199	$\pm$	0.560	 \\	\hline
WD J124825.42$+$550832.61	 & 	192.1050902	 & 	55.1414257	 & 		 & 	17.0714	$\pm$	0.2162	 & 	-107.805	$\pm$	0.217	 & 	-217.504	$\pm$	0.211	 \\	
WD J124825.59$+$550830.91	 & 	192.1057717	 & 	55.1409618	 & 	2.181	 & 	16.9965	$\pm$	0.3220	 & 	-111.970	$\pm$	0.328	 & 	-215.297	$\pm$	0.321	 \\	\hline
WD J132152.71$-$504407.04	 & 	200.4695123	 & 	-50.7351866	 & 		 & 	11.5905	$\pm$	0.3779	 & 	-19.383	$\pm$	0.469	 & 	23.848	$\pm$	0.378	 \\	
WD J132153.92$-$504354.57	 & 	200.4745356	 & 	-50.7317177	 & 	16.940	 & 	11.5753	$\pm$	0.2937	 & 	-17.006	$\pm$	0.355	 & 	24.615	$\pm$	0.280	 \\	\hline
WD J133552.03$+$123709.07	 & 	203.9662259	 & 	12.6193339	 & 		 & 	10.0259	$\pm$	0.0545	 & 	-123.526	$\pm$	0.066	 & 	33.491	$\pm$	0.043	 \\	
WD J133552.15$+$123716.40	 & 	203.9667104	 & 	12.6213767	 & 	7.549	 & 	10.0910	$\pm$	0.3597	 & 	-124.901	$\pm$	0.434	 & 	34.450	$\pm$	0.295	 \\	\hline
WD J140248.74$+$154433.22	 & 	210.7028894	 & 	15.7419024	 & 		 & 	11.0134	$\pm$	0.5616	 & 	-42.396	$\pm$	0.603	 & 	-148.719	$\pm$	0.450	 \\	
WD J140249.85$+$154453.37	 & 	210.7075265	 & 	15.7474894	 & 	25.743	 & 	10.6276	$\pm$	0.6432	 & 	-42.055	$\pm$	0.733	 & 	-149.500	$\pm$	0.544	 \\	\hline
WD J144920.75$+$205419.74	 & 	222.3358875	 & 	20.9054711	 & 		 & 	13.2149	$\pm$	0.1755	 & 	-120.029	$\pm$	0.194	 & 	-3.471	$\pm$	0.193	 \\	
WD J144921.38$+$205421.86	 & 	222.3385018	 & 	20.9060554	 & 	9.040	 & 	13.3466	$\pm$	0.2241	 & 	-123.303	$\pm$	0.236	 & 	-3.649	$\pm$	0.234	 \\	\hline
WD J145314.54$-$225858.51	 & 	223.3095891	 & 	-22.9830958	 & 		 & 	13.8344	$\pm$	0.3679	 & 	-202.568	$\pm$	0.434	 & 	-40.774	$\pm$	0.423	 \\	
WD J145314.72$-$225854.75	 & 	223.3103489	 & 	-22.9820457	 & 	4.542	 & 	14.5914	$\pm$	0.6728	 & 	-206.104	$\pm$	0.687	 & 	-38.284	$\pm$	0.683	 \\	\hline
WD J145739.27$+$272814.87	 & 	224.4131763	 & 	27.4708850	 & 		 & 	15.0460	$\pm$	0.1846	 & 	-92.655	$\pm$	0.167	 & 	18.949	$\pm$	0.183	 \\	
WD J145739.77$+$272820.63	 & 	224.4152345	 & 	27.4724957	 & 	8.766	 & 	15.2571	$\pm$	0.2133	 & 	-92.951	$\pm$	0.196	 & 	22.434	$\pm$	0.208	 \\	\hline
WD J151715.46$-$021902.80	 & 	229.3142711	 & 	-2.3178085	 & 		 & 	10.0298	$\pm$	0.4878	 & 	-31.685	$\pm$	0.505	 & 	-81.818	$\pm$	0.419	 \\	
WD J151715.49$-$021900.84	 & 	229.3143946	 & 	-2.3172395	 & 	2.096	 & 	10.3005	$\pm$	0.4067	 & 	-33.610	$\pm$	0.466	 & 	-76.820	$\pm$	0.373	 \\	\hline
WD J152229.86$-$124741.75	 & 	230.6246930	 & 	-12.7948691	 & 		 & 	12.3186	$\pm$	0.2333	 & 	61.212	$\pm$	0.275	 & 	13.413	$\pm$	0.204	 \\	
WD J152232.31$-$124729.90	 & 	230.6349173	 & 	-12.7915772	 & 	37.800	 & 	12.3110	$\pm$	0.3460	 & 	60.289	$\pm$	0.425	 & 	13.704	$\pm$	0.300	 \\	\hline
WD J152239.94$+$431306.63	 & 	230.6660700	 & 	43.2182436	 & 		 & 	10.9025	$\pm$	0.2353	 & 	-53.525	$\pm$	0.237	 & 	-59.453	$\pm$	0.277	 \\	
WD J152240.18$+$431306.13	 & 	230.6670755	 & 	43.2181088	 & 	2.682	 & 	10.9107	$\pm$	0.2780	 & 	-57.071	$\pm$	0.296	 & 	-58.424	$\pm$	0.341	 \\	\hline
WD J153019.54$-$342353.78	 & 	232.5821497	 & 	-34.3983934	 & 		 & 	13.7398	$\pm$	0.3004	 & 	133.588	$\pm$	0.269	 & 	-27.817	$\pm$	0.214	 \\	
WD J153019.71$-$342356.51	 & 	232.5828774	 & 	-34.3991303	 & 	3.422	 & 	13.2780	$\pm$	0.3026	 & 	136.390	$\pm$	0.281	 & 	-22.085	$\pm$	0.221	 \\	\hline
WD J153034.07$-$572421.42	 & 	232.6400901	 & 	-57.4055468	 & 		 & 	24.4933	$\pm$	0.1103	 & 	-225.415	$\pm$	0.110	 & 	90.537	$\pm$	0.111	 \\	
WD J153034.35$-$572425.84	 & 	232.6413400	 & 	-57.4067441	 & 	4.945	 & 	24.5087	$\pm$	0.1122	 & 	-215.592	$\pm$	0.113	 & 	97.269	$\pm$	0.114	 \\	\hline
WD J155612.11$+$182907.51	 & 	239.0495810	 & 	18.4854246	 & 		 & 	11.9042	$\pm$	0.5971	 & 	-187.152	$\pm$	0.549	 & 	0.957	$\pm$	0.544	 \\	
WD J155612.11$+$182909.89	 & 	239.0496053	 & 	18.4860857	 & 	2.381	 & 	11.5064	$\pm$	0.6388	 & 	-179.427	$\pm$	0.566	 & 	1.689	$\pm$	0.578	 \\	\hline
WD J155755.23$-$383242.75	 & 	239.4797304	 & 	-38.5458134	 & 		 & 	22.4894	$\pm$	0.1755	 & 	-71.914	$\pm$	0.197	 & 	-136.273	$\pm$	0.165	 \\	
WD J155755.87$-$383243.60	 & 	239.4824072	 & 	-38.5460503	 & 	7.585	 & 	22.2701	$\pm$	0.1314	 & 	-69.982	$\pm$	0.148	 & 	-136.232	$\pm$	0.122	 \\	\hline
WD J155925.30$-$374421.23	 & 	239.8554544	 & 	-37.7391697	 & 		 & 	10.6021	$\pm$	0.1033	 & 	6.161	$\pm$	0.125	 & 	14.041	$\pm$	0.089	 \\	
WD J155925.47$-$374420.84	 & 	239.8561420	 & 	-37.7390513	 & 	2.003	 & 	10.4746	$\pm$	0.3218	 & 	5.720	$\pm$	0.394	 & 	16.134	$\pm$	0.279	 \\	\hline
WD J155943.13$+$181714.22	 & 	239.9291537	 & 	18.2870074	 & 		 & 	11.9905	$\pm$	0.2996	 & 	-117.872	$\pm$	0.282	 & 	-63.075	$\pm$	0.243	 \\	
WD J155943.53$+$181716.10	 & 	239.9308003	 & 	18.2875209	 & 	5.924	 & 	12.1533	$\pm$	0.3741	 & 	-118.682	$\pm$	0.355	 & 	-63.816	$\pm$	0.305	 \\	\hline
WD J164055.84$+$034533.83	 & 	250.2318826	 & 	3.7594146	 & 		 & 	21.6318	$\pm$	0.1844	 & 	-175.533	$\pm$	0.229	 & 	3.558	$\pm$	0.175	 \\	
WD J164055.96$+$034534.08	 & 	250.2324524	 & 	3.7594652	 & 	2.055	 & 	21.8072	$\pm$	0.1903	 & 	-164.487	$\pm$	0.234	 & 	-0.482	$\pm$	0.179	 \\	\hline
WD J164837.72$-$353818.39	 & 	252.1571658	 & 	-35.6386258	 & 		 & 	16.7820	$\pm$	0.0406	 & 	1.186	$\pm$	0.054	 & 	-41.003	$\pm$	0.041	 \\	
WD J164837.72$-$353821.08	 & 	252.1571717	 & 	-35.6393620	 & 	2.650	 & 	16.0062	$\pm$	0.2234	 & 	3.575	$\pm$	0.282	 & 	-39.068	$\pm$	0.218	 \\	\hline
WD J164853.70$+$494858.11	 & 	252.2233167	 & 	49.8159809	 & 		 & 	14.6919	$\pm$	0.1851	 & 	-63.683	$\pm$	0.211	 & 	-36.345	$\pm$	0.268	 \\	
WD J164854.12$+$494852.65	 & 	252.2250758	 & 	49.8144713	 & 	6.799	 & 	15.0175	$\pm$	0.1746	 & 	-64.438	$\pm$	0.204	 & 	-35.085	$\pm$	0.254	 \\	\hline
WD J171937.19$+$013407.20	 & 	259.9044771	 & 	1.5680558	 & 		 & 	15.4507	$\pm$	0.3276	 & 	-108.509	$\pm$	0.273	 & 	-137.606	$\pm$	0.198	 \\	
WD J171937.68$+$013406.90	 & 	259.9065454	 & 	1.5679802	 & 	7.448	 & 	15.9414	$\pm$	0.2917	 & 	-107.030	$\pm$	0.245	 & 	-136.612	$\pm$	0.177	 \\	\hline
WD J172603.91$+$323026.47	 & 	261.5158539	 & 	32.5071464	 & 		 & 	10.4958	$\pm$	0.3264	 & 	-83.315	$\pm$	0.354	 & 	-45.714	$\pm$	0.453	 \\	
WD J172604.97$+$323023.63	 & 	261.5202537	 & 	32.5063565	 & 	13.657	 & 	12.2843	$\pm$	0.3185	 & 	-84.686	$\pm$	0.349	 & 	-46.369	$\pm$	0.363	 \\	\hline
WD J172830.68$+$072114.13	 & 	262.1274743	 & 	7.3526572	 & 		 & 	21.2759	$\pm$	0.1506	 & 	-83.738	$\pm$	0.144	 & 	-285.159	$\pm$	0.114	 \\	
WD J172830.82$+$072115.04	 & 	262.1280467	 & 	7.3528659	 & 	2.177	 & 	21.1428	$\pm$	0.1205	 & 	-80.251	$\pm$	0.116	 & 	-295.357	$\pm$	0.093	 \\	\hline
WD J172929.26$+$291609.73	 & 	262.3711812	 & 	29.2684737	 & 		 & 	24.4408	$\pm$	0.0464	 & 	-146.014	$\pm$	0.047	 & 	-201.389	$\pm$	0.053	 \\	
WD J172929.64$+$291554.77	 & 	262.3727428	 & 	29.2643208	 & 	15.734	 & 	24.3286	$\pm$	0.1034	 & 	-150.217	$\pm$	0.106	 & 	-200.864	$\pm$	0.121	 \\	\hline
WD J173423.57$-$220802.85	 & 	263.5981574	 & 	-22.1341407	 & 		 & 	10.3377	$\pm$	2.0529	 & 	-6.858	$\pm$	1.158	 & 	-3.552	$\pm$	0.805	 \\	
WD J173424.35$-$220807.89	 & 	263.6014344	 & 	-22.1355442	 & 	12.039	 & 	11.5769	$\pm$	2.0196	 & 	-6.785	$\pm$	1.877	 & 	-4.238	$\pm$	1.037	 \\	\hline
WD J180013.41$+$093221.36	 & 	270.0542131	 & 	9.5386208	 & 		 & 	13.4562	$\pm$	0.5311	 & 	-367.027	$\pm$	0.510	 & 	-145.023	$\pm$	0.515	 \\	
WD J180014.15$+$093219.61	 & 	270.0572815	 & 	9.5381441	 & 	11.028	 & 	13.4576	$\pm$	0.3541	 & 	-367.995	$\pm$	0.347	 & 	-142.765	$\pm$	0.336	 \\	\hline
WD J180452.09$-$661706.00	 & 	271.2174211	 & 	-66.2849851	 & 		 & 	12.2625	$\pm$	0.3173	 & 	34.224	$\pm$	0.259	 & 	3.167	$\pm$	0.285	 \\	
WD J180452.87$-$661709.45	 & 	271.2206909	 & 	-66.2859583	 & 	5.890	 & 	11.7501	$\pm$	0.2487	 & 	36.597	$\pm$	0.197	 & 	0.242	$\pm$	0.217	 \\	\hline
WD J182518.67$+$553847.70	 & 	276.3265512	 & 	55.6458787	 & 		 & 	20.8992	$\pm$	0.1391	 & 	-155.196	$\pm$	0.157	 & 	-158.702	$\pm$	0.165	 \\	
WD J182519.30$+$553849.16	 & 	276.3291734	 & 	55.6462749	 & 	5.515	 & 	21.1345	$\pm$	0.1583	 & 	-156.994	$\pm$	0.187	 & 	-161.023	$\pm$	0.188	 \\	\hline
WD J184905.83$-$545424.77	 & 	282.2736209	 & 	-54.9069843	 & 		 & 	16.4783	$\pm$	0.0719	 & 	-87.492	$\pm$	0.053	 & 	-23.537	$\pm$	0.042	 \\	
WD J184906.26$-$545424.55	 & 	282.2753901	 & 	-54.9069503	 & 	3.664	 & 	16.2976	$\pm$	0.1955	 & 	-90.245	$\pm$	0.142	 & 	-29.506	$\pm$	0.113	 \\	\hline
WD J185933.28$-$552909.16	 & 	284.8860454	 & 	-55.4856619	 & 		 & 	15.9770	$\pm$	0.1457	 & 	-332.412	$\pm$	0.122	 & 	48.455	$\pm$	0.111	 \\	
WD J185934.08$-$552911.02	 & 	284.8893795	 & 	-55.4861625	 & 	7.036	 & 	16.0679	$\pm$	0.2934	 & 	-336.312	$\pm$	0.258	 & 	51.426	$\pm$	0.224	 \\	\hline
WD J192129.72$+$091722.37	 & 	290.3734550	 & 	9.2893088	 & 		 & 	13.4110	$\pm$	0.3544	 & 	-88.228	$\pm$	0.341	 & 	-53.618	$\pm$	0.290	 \\	
WD J192129.77$+$091723.68	 & 	290.3736263	 & 	9.2896549	 & 	1.387	 & 	12.6388	$\pm$	0.3536	 & 	-89.969	$\pm$	0.324	 & 	-57.597	$\pm$	0.276	 \\	\hline
WD J192441.31$-$205253.98	 & 	291.1718042	 & 	-20.8816587	 & 		 & 	11.8223	$\pm$	0.2035	 & 	-66.810	$\pm$	0.222	 & 	0.306	$\pm$	0.193	 \\	
WD J192442.27$-$205239.40	 & 	291.1758042	 & 	-20.8776067	 & 	19.845	 & 	11.6835	$\pm$	0.3562	 & 	-67.049	$\pm$	0.422	 & 	1.441	$\pm$	0.348	 \\	\hline
WD J192911.25$-$531325.98	 & 	292.2971418	 & 	-53.2256090	 & 		 & 	18.7678	$\pm$	0.1499	 & 	33.422	$\pm$	0.120	 & 	-388.595	$\pm$	0.122	 \\	
WD J192912.69$-$531352.30	 & 	292.3031317	 & 	-53.2329200	 & 	29.315	 & 	18.3224	$\pm$	0.1461	 & 	33.247	$\pm$	0.113	 & 	-388.460	$\pm$	0.108	 \\	\hline
WD J193230.35$+$401039.60	 & 	293.1256151	 & 	40.1767345	 & 		 & 	12.0382	$\pm$	0.1951	 & 	-143.537	$\pm$	0.209	 & 	-209.070	$\pm$	0.288	 \\	
WD J193231.37$+$401047.67	 & 	293.1298404	 & 	40.1789774	 & 	14.152	 & 	12.1740	$\pm$	0.1987	 & 	-145.526	$\pm$	0.219	 & 	-209.414	$\pm$	0.257	 \\	\hline
WD J194231.45$-$051201.13	 & 	295.6311256	 & 	-5.2005197	 & 		 & 	10.2222	$\pm$	0.2243	 & 	14.337	$\pm$	0.267	 & 	-46.236	$\pm$	0.203	 \\	
WD J194231.65$-$051202.98	 & 	295.6319363	 & 	-5.2010312	 & 	3.441	 & 	10.2913	$\pm$	0.4702	 & 	12.907	$\pm$	0.569	 & 	-45.259	$\pm$	0.434	 \\	\hline
WD J194432.04$-$042459.67	 & 	296.1337793	 & 	-4.4171658	 & 		 & 	10.0265	$\pm$	0.4723	 & 	56.908	$\pm$	0.615	 & 	-133.780	$\pm$	0.309	 \\	
WD J194432.21$-$042501.68	 & 	296.1344541	 & 	-4.4177335	 & 	3.169	 & 	10.4764	$\pm$	0.5755	 & 	56.846	$\pm$	0.780	 & 	-135.002	$\pm$	0.395	 \\	\hline
WD J194530.33$+$465006.84	 & 	296.3734339	 & 	46.8334481	 & 		 & 	34.7726	$\pm$	0.0577	 & 	-455.318	$\pm$	0.066	 & 	-401.491	$\pm$	0.075	 \\	
WD J194530.35$+$465015.52	 & 	296.3735335	 & 	46.8358461	 & 	8.636	 & 	34.7116	$\pm$	0.0509	 & 	-453.184	$\pm$	0.060	 & 	-404.742	$\pm$	0.066	 \\	\hline
WD J201939.74$+$074024.33	 & 	304.9151122	 & 	7.6732672	 & 		 & 	12.6642	$\pm$	0.3726	 & 	-102.335	$\pm$	0.420	 & 	-35.759	$\pm$	0.309	 \\	
WD J201940.16$+$074029.04	 & 	304.9168980	 & 	7.6745669	 & 	7.905	 & 	12.2056	$\pm$	0.3094	 & 	-99.273	$\pm$	0.360	 & 	-37.512	$\pm$	0.279	 \\	\hline
WD J203219.82$+$360854.52	 & 	308.0828312	 & 	36.1482258	 & 		 & 	11.1764	$\pm$	0.1683	 & 	42.475	$\pm$	0.167	 & 	-55.758	$\pm$	0.196	 \\	
WD J203220.71$+$360847.96	 & 	308.0865251	 & 	36.1464066	 & 	12.578	 & 	10.9295	$\pm$	0.2726	 & 	42.554	$\pm$	0.258	 & 	-55.873	$\pm$	0.305	 \\	\hline
WD J203801.72$-$343244.64	 & 	309.5072508	 & 	-34.5465006	 & 		 & 	10.0891	$\pm$	0.3269	 & 	13.427	$\pm$	0.349	 & 	-172.628	$\pm$	0.256	 \\	
WD J203801.84$-$343246.61	 & 	309.5077340	 & 	-34.5470526	 & 	2.450	 & 	11.0761	$\pm$	0.4066	 & 	12.947	$\pm$	0.438	 & 	-174.387	$\pm$	0.333	 \\	\hline
WD J203834.39$+$491650.27	 & 	309.6427864	 & 	49.2805151	 & 		 & 	10.9860	$\pm$	0.2307	 & 	-74.948	$\pm$	0.262	 & 	-26.305	$\pm$	0.297	 \\	
WD J203835.87$+$491625.89	 & 	309.6489571	 & 	49.2737374	 & 	28.379	 & 	11.2517	$\pm$	0.2411	 & 	-73.294	$\pm$	0.276	 & 	-27.176	$\pm$	0.286	 \\	\hline
WD J211534.14$+$324917.86	 & 	318.8935163	 & 	32.8221559	 & 		 & 	10.7101	$\pm$	0.4716	 & 	241.086	$\pm$	0.359	 & 	119.607	$\pm$	0.380	 \\	
WD J211534.59$+$324919.65	 & 	318.8953830	 & 	32.8226572	 & 	5.928	 & 	11.1115	$\pm$	0.4195	 & 	239.708	$\pm$	0.302	 & 	119.794	$\pm$	0.365	 \\	\hline
WD J211723.80$-$415645.46	 & 	319.3522537	 & 	-41.9473075	 & 		 & 	16.8279	$\pm$	0.2305	 & 	516.386	$\pm$	0.220	 & 	-302.708	$\pm$	0.204	 \\	
WD J211724.10$-$415600.61	 & 	319.3534952	 & 	-41.9348531	 & 	44.959	 & 	16.7504	$\pm$	0.2770	 & 	516.339	$\pm$	0.251	 & 	-304.050	$\pm$	0.243	 \\	\hline
WD J211812.19$-$520647.64	 & 	319.5508273	 & 	-52.1134726	 & 		 & 	20.0474	$\pm$	0.1102	 & 	4.279	$\pm$	0.090	 & 	-53.985	$\pm$	0.090	 \\	
WD J211812.39$-$520646.74	 & 	319.5517120	 & 	-52.1132481	 & 	2.116	 & 	20.1444	$\pm$	0.1618	 & 	12.135	$\pm$	0.132	 & 	-59.358	$\pm$	0.146	 \\	\hline
WD J212436.88$-$301446.41	 & 	321.1542645	 & 	-30.2467007	 & 		 & 	12.6561	$\pm$	0.5276	 & 	120.055	$\pm$	0.551	 & 	-106.110	$\pm$	0.487	 \\	
WD J212437.00$-$301443.03	 & 	321.1548038	 & 	-30.2457645	 & 	3.764	 & 	12.0924	$\pm$	0.4744	 & 	121.762	$\pm$	0.481	 & 	-107.369	$\pm$	0.432	 \\	\hline
WD J214538.15$+$110627.20	 & 	326.4097575	 & 	11.1059347	 & 		 & 	14.9158	$\pm$	0.6026	 & 	180.090	$\pm$	0.730	 & 	-364.789	$\pm$	0.601	 \\	
WD J214538.58$+$110619.69	 & 	326.4115802	 & 	11.1038357	 & 	9.928	 & 	13.8814	$\pm$	0.4204	 & 	181.814	$\pm$	0.540	 & 	-366.779	$\pm$	0.424	 \\	\hline
WD J215817.63$-$214103.59	 & 	329.5732982	 & 	-21.6843195	 & 		 & 	12.9907	$\pm$	0.3795	 & 	-37.834	$\pm$	0.423	 & 	2.371	$\pm$	0.370	 \\	
WD J215817.89$-$214108.81	 & 	329.5743465	 & 	-21.6857653	 & 	6.276	 & 	12.9103	$\pm$	0.3943	 & 	-39.952	$\pm$	0.452	 & 	3.614	$\pm$	0.394	 \\	\hline
WD J222521.42$-$164000.16	 & 	336.3396502	 & 	-16.6672945	 & 		 & 	11.1577	$\pm$	0.6504	 & 	84.103	$\pm$	0.674	 & 	-131.735	$\pm$	0.543	 \\	
WD J222522.76$-$164039.47	 & 	336.3452289	 & 	-16.6782316	 & 	43.823	 & 	11.1115	$\pm$	0.6583	 & 	84.101	$\pm$	0.746	 & 	-133.527	$\pm$	0.562	 \\	\hline
WD J222542.67$-$011405.31	 & 	336.4273234	 & 	-1.2349274	 & 		 & 	13.3670	$\pm$	0.1650	 & 	-100.930	$\pm$	0.165	 & 	-26.960	$\pm$	0.157	 \\	
WD J222543.52$-$011359.51	 & 	336.4308864	 & 	-1.2333078	 & 	14.087	 & 	12.9917	$\pm$	0.3361	 & 	-100.944	$\pm$	0.363	 & 	-23.515	$\pm$	0.351	 \\	\hline
WD J223755.87$-$224712.84	 & 	339.4834022	 & 	-22.7875926	 & 		 & 	14.8839	$\pm$	0.2004	 & 	124.775	$\pm$	0.192	 & 	-155.773	$\pm$	0.190	 \\	
WD J223756.73$-$224715.01	 & 	339.4869566	 & 	-22.7882015	 & 	11.999	 & 	15.1064	$\pm$	0.1904	 & 	124.352	$\pm$	0.184	 & 	-157.346	$\pm$	0.181	 \\	\hline
WD J224827.95$-$583031.01	 & 	342.1161408	 & 	-58.5094421	 & 		 & 	19.0166	$\pm$	0.1092	 & 	-36.568	$\pm$	0.088	 & 	-186.730	$\pm$	0.124	 \\	
WD J224829.50$-$583036.39	 & 	342.1225713	 & 	-58.5109334	 & 	13.230	 & 	19.1171	$\pm$	0.1289	 & 	-38.476	$\pm$	0.103	 & 	-185.783	$\pm$	0.144	 \\	\hline
WD J230418.93$-$070124.51	 & 	346.0799701	 & 	-7.0244716	 & 		 & 	18.2338	$\pm$	0.0927	 & 	240.421	$\pm$	0.097	 & 	-224.484	$\pm$	0.080	 \\	
WD J230419.49$-$070149.84	 & 	346.0823042	 & 	-7.0315090	 & 	26.672	 & 	18.5819	$\pm$	0.3101	 & 	240.012	$\pm$	0.345	 & 	-225.269	$\pm$	0.281	 \\	\hline
WD J230959.10$+$301111.01	 & 	347.4968786	 & 	30.1861831	 & 		 & 	10.8602	$\pm$	0.3961	 & 	122.004	$\pm$	0.311	 & 	-44.872	$\pm$	0.329	 \\	
WD J230959.14$+$301107.20	 & 	347.4970179	 & 	30.1851302	 & 	3.815	 & 	11.1141	$\pm$	0.3385	 & 	119.535	$\pm$	0.275	 & 	-46.748	$\pm$	0.275	 \\	\hline
WD J232115.34$+$010211.96	 & 	350.3134624	 & 	1.0355488	 & 		 & 	16.5661	$\pm$	0.2698	 & 	-104.998	$\pm$	0.267	 & 	-248.714	$\pm$	0.289	 \\	
WD J232115.70$+$010224.59	 & 	350.3149464	 & 	1.0390470	 & 	13.679	 & 	16.3779	$\pm$	0.2954	 & 	-104.747	$\pm$	0.270	 & 	-251.384	$\pm$	0.285	 \\	\hline
WD J235427.59$-$711909.32	 & 	358.6194068	 & 	-71.3203031	 & 		 & 	10.1262	$\pm$	0.3332	 & 	318.796	$\pm$	0.411	 & 	-234.997	$\pm$	0.474	 \\	
WD J235428.19$-$711907.67	 & 	358.6219189	 & 	-71.3198456	 & 	3.332	 & 	11.0856	$\pm$	0.2945	 & 	319.537	$\pm$	0.313	 & 	-234.464	$\pm$	0.390	 \\	
\enddata
\end{deluxetable*}

\clearpage
\bibliography{references}{}
\bibliographystyle{aasjournal}

\end{document}